\documentclass[prc,aps,floats,onecolumn,floatfix,showpacs,nofootinbib,%
superscriptaddress]{revtex4}

\usepackage{amsmath}
\usepackage{amssymb}
\usepackage{amsbsy}
\usepackage{amsfonts}
\usepackage{graphicx}
\usepackage{tabularx}
\usepackage{multirow}
\usepackage{lscape}
\usepackage{rotating}
\usepackage{pstricks}
\usepackage{feynmf}
\usepackage{feynmf}
\usepackage{hyperref}
\usepackage{orcidlink}
\usepackage{tikz}
\usetikzlibrary{shapes.geometric, arrows}
\tikzstyle{startstop} = [rectangle, rounded corners, minimum width=3cm, minimum height=1cm,text centered, draw=black, fill=red!30]
\tikzstyle{io} = [trapezium, trapezium left angle=70, trapezium right angle=110, minimum width=3cm, minimum height=1cm, text centered, draw=black, fill=blue!30]
\tikzstyle{process} = [rectangle, minimum width=3cm, minimum height=1cm, text centered, draw=black, fill=orange!30]
\tikzstyle{decision} = [diamond, minimum width=3cm, minimum height=1cm, text centered, draw=black, fill=green!30]
\tikzstyle{arrow} = [thick,->,>=stealth]

\begin{document}

\title{Nuclear surface energy in a semiclassical
Extended Thomas-Fermi approach with finite-range interactions}


\author{D. Davesne}
\email{davesne@ipnl.in2p3.fr}
\affiliation{Institut de Physique des 2 infinis de Lyon, CNRS-IN2P3, UMR 5822, Universit\'e Lyon 1 \\
             43 Bd. du 11 Novembre 1918, F-69622 Villeurbanne cedex, France}

\author{Y. Lallouet\,\orcidlink{9-0002-2029-53630}}
\affiliation{Lycée Malherbe, 14, Avenue Albert Sorel, 14000 Caen, France}

\author{A. Pastore\,\orcidlink{0000-0003-3354-6432}}
\email{alessandro.pastore@cea.fr}
\affiliation{ CEA, DES, IRESNE, DER, SPRC, F-13108 Saint Paul Lez Durance, France}

\author{J. Navarro}
\email{navarro@ific.uv.es}
\affiliation{IFIC (CSIC-Universidad de Valencia), Parque Cient\'{\i}fico, Catedr\'atico Jos\'e Beltr\'an 2, E-46.980-Paterna, Spain}

\author{X. Vi\~nas}
\email{xavier@fqa.ub.edu}
\affiliation{Departament de F\'isica Qu\`antica i Astrof\'isica (FQA),
Universitat de Barcelona (UB), Mart\'i i Franqu\`es 1, E-08028 Barcelona, Spain}
\affiliation{Institut de Ci\`encies del Cosmos (ICCUB),
Universitat de Barcelona (UB), Mart\'i i Franqu\`es 1, E-08028 Barcelona, Spain}
\affiliation{Institut Menorqu\'i d'Estudis, Cam\'i des Castell 28, 07702 Ma\'o, Spain}

\begin{abstract}
Using the Gogny finite-range interaction, we investigate a series of semiclassical approximations to the Fock term entering the calculation of the nuclear surface energy. By comparing these approximations with the exact results obtained from the full Hartree–Fock solution in semi-infinite nuclear matter, we derive a simple pocket formula that can be incorporated into fitting protocols to estimate the surface energy coefficient with excellent accuracy.
\end{abstract}

 
\date{\today}


\maketitle

\section{Introduction}

In order to determine the coupling constants of an effective nucleon--nucleon (NN) interaction for mean-field applications~\cite{ring:book}, it is essential to define an appropriate optimisation procedure~\cite{kor10,kor12,kor14}. Typically, this involves minimising a penalty function~\cite{dob14} constructed from a combination of finite-nucleus observables, such as masses, charge radii, single-particle splittings, \dots together with properties of infinite nuclear matter (INM), including the binding energy per particle, saturation density, incompressibility coefficient, symmetry energy, etc.~\cite{was12}.

One advantage of including pseudo-observables from INM is that they can be related to other nuclear quantities. For instance, one may mention the connection between the nuclear incompressibility of INM and the excitation energy of giant monopole resonances in finite nuclei~\cite{gar18}, the correlation between the centroid energy of the isovector giant quadrupole resonance and the nuclear effective mass~\cite{boh79}, or the relationship between the neutron-skin thickness of $^{208}$Pb and the slope of the symmetry energy~\cite{Bro00,cen09}.
Although recent work~\cite{atk20} has raised some doubts about the robustness of the links between INM properties and finite-nucleus observables, INM remains a valuable guide in optimisation procedures. Indeed, simple calculations performed in infinite matter can provide reasonable initial estimates for the interaction parameters, which can subsequently be refined using experimental data from finite nuclei.

Contrary to infinite nuclear matter, finite nuclei possess a surface whose associated energy plays a pivotal role in determining important physical properties, such as deformation~\cite{bar82,rys19}, fusion mechanisms~\cite{ste20,sal11,dut10,gha15,gol13}, fission-barrier heights~\cite{tos13,gou05}, and cluster decay~\cite{raj17}. To extract the surface energy without contamination from finite-size effects, one can perform a Hartree--Fock (HF) calculation in semi-infinite nuclear matter (SINM) for a given NN interaction.
Following the pioneering works of Swiatecki~\cite{swi51,mye66}, SINM is an idealised system that is translationally invariant along the two infinite directions, $x$ and $y$, and bounded by a surface perpendicular to the $z$ axis. The neutron and proton density profiles are denoted by $\rho_q(z)$ ($q=n,p$). Since we are interested only in the surface energy of symmetric matter, we consider exclusively the isoscalar density, $\rho(z)=\rho_n(z)+\rho_p(z)$. Deep inside the matter, \emph{i.e.} for $z\rightarrow -\infty$, $\rho(z)\rightarrow\rho_0$, where $\rho_0$ is the saturation density of symmetric infinite nuclear matter, while ${\cal E}(z)/\rho(z)\rightarrow a_v$, the corresponding energy per particle. Outside the matter, \emph{i.e.} for $z\rightarrow\infty$, $\rho(z)\rightarrow0$ and ${\cal E}(z)/\rho(z)\rightarrow0$.

The surface energy per unit area is given by the difference between the energies per unit area of SINM and INM systems. Denoting by ${\cal E}[\rho(z)]$ the energy density  one has
\begin{eqnarray}\label{eq:EsurS}
    \frac{E}{S} = \int_{-\infty}^{\infty} {\rm d}z \, \left[ \mathcal{E}(z)-a_v\rho(z)\right] \equiv  \int_{-\infty}^{+\infty} {\rm d}z \, \left[ \mathcal{E}[\rho] - \frac{\rho}{\rho_0} \mathcal{E}[\rho_0] \right]. 
\end{eqnarray}
To get the surface energy per nucleon, or surface energy coefficient $a_s$, we multiply by the elementary surface $4 \pi r_0^2$, where $r_0$ is unit radius, defined through the condition $\frac{4}{3} \pi r_0^3 \rho_0 = 1$
\begin{eqnarray}\label{eq:surf}
   a_s = \left(\frac{36 \pi}{\rho_0^2}\right)^{\frac{1}{3}} \frac{E}{S}. 
\end{eqnarray}

In the case of a zero-range interaction such as Skyrme~\cite{sky58}, the HF equations in SINM boil down to a series of second order differential equations that can be handled easily using for instance the Numerov algorithm~\cite{nou24}. See Ref.~\cite{jod16} for more details.
In the case of finite-range interactions, such as the Gogny interaction~\cite{dec80}, the problem reduces to solving an integro-differential Schr\"odinger equation with appropriate asymptotic boundary conditions~\cite{cot78,dav23}. Although feasible, such calculations are too computationally demanding to be performed repeatedly during an optimisation procedure, even when using the recently developed, faster, and more efficient method based on the Lagrange mesh~\cite{dav25b,bay15}. The resulting surface-energy coefficients are listed in Table~\ref{tab:HF} for several currently used Gogny parameterisations. The calculations were performed using a finite slab of SINM, and the quoted uncertainties arise from averaging over different box sizes.

\begin{table}[!h]
    \centering
    \begin{tabular}{l c|c|c}
        \hline 
        \hline
         & & $a_s$[MeV]  &$a_s$[MeV] (no s.o.)  \\
        \hline
        D1  &\cite{dec80}     &19.74 $\pm$ 0.02   & 20.92 $\pm$ 0.02\\
        D1S& \cite{dec80}      & 17.91 $\pm$ 0.01  &19.35 $\pm$ 0.02\\
        D1N &\cite{chap08}      &18.04 $\pm$ 0.01   & 19.29 $\pm$ 0.02\\
        D1M &\cite{gor09d1m}     &18.13 $\pm$ 0.01  & 19.44 $\pm$ 0.01\\
        D1M* &\cite{gon18}   &18.20 $\pm$ 0.01  & 19.60 $\pm$ 0.01  \\
        D1MK  &\cite{tag22} &18.26 $\pm$ 0.03 & 19.59 $\pm$ 0.02\\
        D3G3 &\cite{bat23}    &17.83 $\pm$ 0.02  & 19.08 $\pm$ 0.02  \\
        D3G3M & \cite{bat24}&18.37 $\pm$ 0.02  & 19.84 $\pm$ 0.02  \\
        D250 &\cite{bla95}& 18.61 $\pm$ 0.02 & 20.07 $\pm$ 0.02\\
        D260 &\cite{bla95}& 19.54 $\pm$ 0.02 & 20.95 $\pm$ 0.02 \\
        D280 &\cite{bla95}& 19.77 $\pm$ 0.02 & 21.08 $\pm$ 0.02 \\
        D300 &\cite{bla95}& 19.58 $\pm$ 0.02 & 21.18 $\pm$ 0.02 \\
         D1P &\cite{far99}     &17.06 $\pm$ 0.02  & 18.97 $\pm$ 0.02 \\
        D1PK  &\cite{tag22} &17.74 $\pm$ 0.01 &19.42 $\pm$ 0.02\\
         \hline
                  \hline
    \end{tabular}
    \caption{The surface energy coefficient obtained from a Hartree-Fock calculation for various Gogny interactions. The last column contains the results obtained removing the spin orbit term. See Ref.~\cite{dav25b} for details.}
    \label{tab:HF}
\end{table}

For a detailed review of properties of the various Gogny interactions we refer the reader to Refs.~\cite{sel14,zie23}. Briefly, starting from the original D1 interaction~\cite{dec80}, successive parameterisations have been developed to address specific shortcomings. For example, D1S was designed to reduce the excessively large surface energy, D1M to improve the equation of state of neutron matter and the description of nuclear masses, while D1N, D1M*, D3G3, and D3G3M were proposed to improve the astrophysical properties of the interaction. The D250--D300 family was  adjusted to investigate the nuclear incompressibility, whereas D1P was developed to improve pairing properties. Finally, we note that two additional interactions, D1MK and D1PK, were proposed in Ref.~\cite{tag22} to better reproduce the neutron-skin thickness of $^{48}$Ca. These two interactions were not included in our previous calculations reported in Ref.~\cite{dav25b}.

It is worth noting that these interactions do not all provide the same level of accuracy in the description of finite nuclei. In particular, some of them, such as D250, D1P, and D1PK, fail to reproduce basic observables such as nuclear binding energies, leading to discrepancies as large as 50 MeV or more for a heavy nucleus such as $^{208}$Pb.

Despite the remarkable improvement achieved in HF calculations for SINM, their computational cost remains prohibitively high for efficient use in parameter-fitting procedures. A possible way to overcome this limitation is to employ semiclassical techniques~\cite{gra79,gra80,bra85,bra97,cen90,cen98,sou00} to accelerate the calculations. These methods have proven to be highly successful for Skyrme interactions. In particular, using a \emph{pocket} formula, it is possible to estimate with good accuracy the surface energy obtained from fully self-consistent Hartree--Fock calculations~\cite{jod16,Proust2023}. However, for finite-range Gogny interactions, the evaluation of the Fock term~\cite{sou95,far99} remains computationally demanding. The aim of the present article is therefore to review the most widely used semiclassical approximations, to calculate with them the SINM surface energy, comparing the resulting predictions with the available quantal calculations reported in Table~\ref{tab:HF}, and to
derive a simple and efficient expression for the surface energy which could be incorporated in the fit protocol of a Gogny interaction. 

The article is organised as follows. In Sec.~\ref{Sec:Fock}, we discuss the structure of the Fock term for a generic Gogny interaction and the various semiclassical approximations to the density matrix. In Sec.~\ref{Sec:SINM}, we investigate the surface energy in semi-infinite nuclear matter, using a simple variational density profile. The contributions of the different interaction terms are explicitely shown. In Sec.~\ref{sec:results}, we present our results based on those semiclassical approximations. In Sec.~\ref{sec:NL}, we quantify the effect of non-local terms on the surface energy using a specific semiclassical approximation to the density matrix. Finally, we present our conclusions in Sec.~\ref{sec:conclusion}.


\section{Approximations to the density matrix}\label{Sec:Fock}

In its \emph{standard} form, the Gogny interaction is written as a sum of three terms~\cite{dec80}
\begin{equation}\label{eq:Gogny:standard}
V(\mathbf{r}_1,\mathbf{r}_2)=V_C(\mathbf{r}_1,\mathbf{r}_2)+V_{DD}(\mathbf{r}_1,\mathbf{r}_2)+V_{SO}(\mathbf{r}_1,\mathbf{r}_2),
\end{equation}
corresponding to the central, density dependent and spin-orbit terms respectively. They read
\begin{eqnarray}
V_C(\mathbf{r}_1,\mathbf{r}_2)&=&\sum_i (W_i+B_iP_\sigma-H_iP_\tau-M_iP_\sigma P_\tau)e^{-r_{12}^2/\mu_i^2}\,,\\
V_{DD}(\mathbf{r}_1,\mathbf{r}_2)&=&t_3(1+x_3P_\sigma) \rho^\gamma \delta(\mathbf{r}_{12})\,,\label{eq:DD}\\
V_{SO}(\mathbf{r}_1,\mathbf{r}_2)&=&i W_0 (\mathbf{k}' \times \mathbf{k})(\sigma_1+\sigma_2) \delta(\mathbf{r}_{12}),
\end{eqnarray}
where ${\bf r_{12}} = {\bf r_1} - {\bf  r_2}$.
Typically the  central term is composed by a sum over two Gaussians for the D1 family and over three for the D3 family. Several parametrisations of this interaction exist in the literature and we refer to Refs.~\cite{sel14,dav25} for an extensive discussion concerning their INM properties. Extensions to Eq.~\eqref{eq:Gogny:standard} have been suggested such as the inclusion of an explicit tensor~\cite{gra13,zie23,zie26}, a finite-range spin-orbit~\cite{zie23,zie26} or a modified density dependent term~\cite{cha15}, but they will not considered here.

As already mentioned, the main difficulty in dealing with finite-range interactions lies in the treatment of the Fock term, since it explicitly depends on the density matrix $\rho({\bf r},{\bf r'})$. In the case of a Gogny interaction, the exchange contribution to the SINM energy reads 
\begin{equation}\label{eq:Exch}
E_E  =  -\frac{1}{8} \sum_i \left( W_i + 2B_i  - 2 H_i - 4 M_i \right) \int d{\bf r} d{\bf r'}  \rho^*({\bf r},{\bf r'}) {\rm e}^{-({\bf r} - {\bf r '})^2 / \mu_i^2} \rho({\bf r},{\bf r'}) \,.
\end{equation}
It is convenient to use the center-of-mass and relative coordinates ${\bf R}=({\bf r}+{\bf r'})/2$ and ${\bf s}={\bf r}-{\bf r'}$, respectively. The full quantal calculation of  $\rho({\bf R},{\bf s})$ for a finite or a semi-infinite system is demanding. To overcome the problem, various semiclassical approaches have been developed in the scientific literature. The simplest one, widely used for infinite range interactions such as Coulomb, 
is the Slater (SL) approximation~\cite{sla51}, for which the non-diagonal part of the density matrix is replaced by its expression in the homogeneous matter as
\begin{eqnarray}\label{eqn:SL}
\rho_{\rm{SL}}({\bf R}, {\bf s}) & = & \rho({\bf R}) \, \hat{\jmath} _1(k_F s),
\end{eqnarray}
where a modified spherical Bessel function has been defined as
\begin{eqnarray}
\hat{\jmath}_{\ell}(x) & = & (2\ell+1)!! \frac{j_{\ell}(x)}{x^{\ell}}. 
\end{eqnarray}
A more refined alternative has been elaborated by Negele and Vautherin  (NV)~\cite{neg72,neg75}. It consists in making an $s$-expansion of the density matrix in such a way that the leading contribution corresponds to the Slater approximation and the corrective terms take into account finite size effects. Up to second order, it is written as 
\begin{equation}
\rho_{\rm{NV}}({\bf R},{\bf s})  =  \rho({\bf R}) \hat{\jmath}_1(k s) +\frac{1}{6} s^2 \hat{\jmath} _3(k s)\left[ \frac{1}{4}\nabla^2 \rho({\bf R}) - \tau({\bf R}) + \frac{3}{5} k^2 \rho({\bf R}) \right] , \label{eqn:NV}
\end{equation}
where $k$ is some average value of the relative momentum of the two interacting particles. Truncating the infinite sum to second order, as above, requires a clever choice for $k$. Negele and Vautherin suggested to replace $k({\bf R})$ by the local Fermi momentum $k_F=\left(3 \pi^2 \rho({\bf R})/2\right)^{1/3}$ so that the second term in Eq.~\eqref{eqn:NV} appears as a correction of the Slater approximation. Another possible choice for $k({\bf R})$  has been suggested by Campi and Bouyssy (CB)~\cite{cam78} 
\begin{eqnarray}
\hat{k}({\bf R}) & = & \left[ \frac{5}{3\rho({\bf R}) } \left( \tau({\bf R}) - \frac{1}{4}\nabla^2 \rho({\bf R})\right) \right]^{1/2}\,. 
\end{eqnarray}
This choice cancels out exactly the second order term of the density matrix expansion. As a consequence, the density matrix takes the familiar form of the Slater approximation, but with  $\hat{k}$ instead of $k_F$ in the argument of $\hat{\jmath} _1$
\begin{eqnarray}\label{eqn:CB}
\rho_{\rm{CB}}({\bf R},{\bf s}) &=& \rho({\bf R}) \hat{\jmath} _1(\hat k s).
\end{eqnarray}
Notice that Eqs.~\eqref{eqn:NV}-\eqref{eqn:CB} involve the kinetic energy density $\tau({\bf R})$. Therefore, an approximation for $\tau$ implies 
an approximation for $\rho({\bf R},{\bf s})$ and consequently an approximation for $E_E$.
Starting from the Wigner-Kirkwood distribution function~\cite{ring:book}, one can deduce the so-called Extended Thomas Fermi (ETF) 
approach~\cite{gra79,gra80,bra85} to the kinetic energy density. Its expression
up to fourth order gradients reads 
\begin{eqnarray}
\tau & = &
\frac{3}{5}\left(3\pi^2\right)^{2/3}\rho^{5/3} + \frac{\Delta \rho}{3} +  \frac{(\nabla \rho)^2}{36 \rho}
 \nonumber \\
& & +\left(3\pi^2\right)^{-2/3} \bigg\{
 \rho^{-2/3} \,\frac{\Delta\Delta \rho}{180}
 + \rho^{-5/3} \left[ - \frac{1}{72} (\nabla \rho \cdot \nabla) \Delta \rho
- \frac{7}{1080} (\Delta \rho)^2 - \frac{7}{2160} \Delta (\nabla \rho)^2 \right]
 \nonumber \\
& &  + \rho^{-8/3} \left[ \frac{7}{324} (\nabla \rho)^2 \Delta \rho
+ \frac{23}{810} (\nabla \rho \cdot \nabla)^2 \rho \right]
- \rho^{-11/3} \, \frac{1}{45} (\nabla \rho)^4
\bigg\}\,.
\label{eq:tau4}
\end{eqnarray}
The first term corresponds to the usual Thomas--Fermi (TF) approximation. The first three terms, which contain at most second-order gradients, define the so-called ETF2 approximation, while the complete expression corresponds to the ETF4 approximation (see~\cite{bra85,pro22} for details). In the following, we adopt the following notation. The labels SL, SL2, and SL4 denote the Slater approximation combined with the Thomas--Fermi TF, ETF2, and ETF4 expressions for the kinetic energy density, respectively. Similarly, the same notation is used for the Negele--Vautherin and Campi--Bouyssy approximations.

\section{Surface energy of semi-infinite nuclear matter}\label{Sec:SINM}

The calculation of $a_s$ requires the use of a density profile $\rho(z)$ in Eq.~\eqref{eq:EsurS}. In general, such a profile is obtained by minimizing the surface energy and solving the resulting  Euler-Lagrange equation for $\rho(z)$. In this article, we shall assume instead a simple Fermi profile for the density
\begin{equation}
\rho(z) = \rho_0 \left( 1 + {\rm e}^{\alpha z} \right) ^{-1}, \label{density:prof:SINM}
\end{equation}
with a single variational parameter $\alpha$. In the following, we shall employ the notations $k_F(z)=(3 \pi^2 \rho(z)/2)^{1/3}$ and  $k_{F0}=(3 \pi^2 \rho_0/2)^{1/3}$ for the Fermi momentum of SINM and INM, respectively, and $x(z)=(1+{\rm e}^{\alpha z})^{-1}$.

It is convenient to treat separately the different contributions of the interaction to the surface energy as a sum of kinetic (T), density dependent (DD), spin orbit (SO), direct (D) and exchange (E) central terms as

\begin{equation}
\frac{E}{S} = \frac{E}{S}\bigg|_T + \frac{E}{S}\bigg|_{DD} + \frac{E}{S}\bigg|_{SO} + \frac{E}{S}\bigg|_D + \frac{E}{S}\bigg|_E\,.
\end{equation}

We present now the expressions of these terms as a function of the variational parameter $\alpha$. 

\subsection{Kinetic energy}\label{subsec:T}
Following Ref.~\cite{gra79}, the kinetic energy contribution up to ETF4 order is written as
\begin{equation}
\frac{E}{S}\bigg|_{T}   =   A \alpha^3 + B \alpha + \frac{C}{\alpha} \, ,
\end{equation}
where   
\begin{eqnarray}
A & = & \frac{\hbar^2}{2 m} \left(\frac{3\pi^2}{2}\right)^{-2/3} \rho_0^{1/3} \frac{1}{840}\label{Energy1},  \\
B & = & \frac{\hbar^2}{2 m} \rho_0 \frac{1}{72}\label{Energy2},  \\
C & = & -\frac{3}{5} \left(\frac{3\pi^2}{2}\right)^{2/3} \rho_0^{5/3} \frac{\hbar^2}{2m} \frac{3}{2} \left[ 1 + \frac{\pi}{3\sqrt{3}} - \ln 3 \right] \label{Energy3}.
\end{eqnarray}
The term $A$ corresponds to ETF4 order, $B$ to ETF2 order, and $C$ to TF order.

\subsection{Density-dependent term} \label{subsec:DD}
The density dependent term is the same as for Skyrme interactions and has already been calculated by example in Ref.~\cite{pro22}. It reads 
\begin{eqnarray}
\frac{E}{S}\bigg|_{DD}  &=& - \, \frac{1}{16} \, 6 t_3 \,  \rho_0^{\gamma+2} C_\gamma \frac{1}{\alpha}\label{Energy:DD:SINM}\,.
\end{eqnarray}
The coefficient $C_\gamma$ is given in Tab.~\ref{tab:dd_power} for different values of $\gamma$ currently used in the literature.
\begin{table}[h!]
\begin{center}
\begin{tabular}{c|cc}
\hline  
\hline
  $\gamma$  && $C_\gamma$ \\ 
\hline
  $1$   && $ \frac{3}{2}$                                               \\[0.3mm]
  $2/3$ && $ \frac{21}{10} + \frac{\pi}{2\sqrt{3}} - \frac{3\log 3}{2}$ \\[0.3mm]
  $1/2$ && $ \frac{8}{3}  - \log 4$                                    \\[0.3mm] 
  $1/3$ && $ \frac{15}{4}  - \frac{\pi}{2\sqrt{3}} - \frac{3\log 3}{2}$  \\[0.3mm]
  $1/4$ && $ \frac{24}{5}  - \frac{\pi}{2}         - \log 8$            \\[0.3mm]
  $1/5$ && $  \frac{35}{6} - \frac{5\log5}{4} 
	                          - \frac{\sqrt{5}}{4} \log \frac{\sqrt{5}+1}{\sqrt{5}-1} 
	                          - \frac{\pi}{2} \sqrt{1+\frac{2}{\sqrt{5}}}$ \\[0.3mm]
  $1/6$ && $ \frac{48}{7}  - \frac{\pi\sqrt{3}}{2} - \frac{3\log3}{2} 
	                          - \log 4$                                    \\[0.3mm]
 $0.92$ &&  1.4679                               \\[0.3mm]
\noalign{\smallskip} \hline 
 \hline 
\end{tabular}
\caption{Explicit expressions of the $C_\gamma$ constant  entering Eq.~\eqref{Energy:DD:SINM} for a variety of $\gamma$ values typically used in the literature for the density dependent terms~\cite{dut12}. }
\label{tab:dd_power}
\end{center}
\end{table}

\subsection{Spin-orbit term}\label{subsec:SO}
Similarly, the spin-orbit term, which is the same used with  Skyrme-type interactions can be found in literature \cite{jod16}. At ETF2 order it reads
\begin{equation}
\frac{E}{S} \bigg|_{SO} = -\frac{3}{64}  W_0^2 \rho_0^3 \frac{m}{\hbar^2} \alpha \,,
\end{equation}
The calculation at ETF4 order was first done by Grammaticos and Voros~\cite{gra80}, including an effective mass. However, their final result contains a few mistakes. Indeed, the coefficient $A$ entering their formula (III.25), related to the $\alpha^3$ term of the surface energy, diverges when the effective mass is equal to the bare mass. Using their notation $b=(f-1)/(\rho/\rho_0)$ and $f=m/m*$, the correct expression for this coefficient is
\begin{eqnarray}
A & = & \left(\frac{3\pi^2}{2}\right)^{-2/3}\left(\frac{m}{\hbar^2}\right)\frac{9W_0^2}{16} \rho_0^{7/3} \left(-\frac{1}{12 b^4}-\frac{13}{168 b^3}+\frac{4}{105 b^2} + \frac{9}{520 b} +\left(\frac{1}{9 b^4}+\frac{1}{6 b^3}-\frac{1}{18 b}\right) D \right) \nonumber \\
& & + \left(\frac{3\pi^2}{2}\right)^{-2/3} \rho_0^{13/3} \left(\frac{m}{\hbar^2}\right)^3\frac{9W_0^4}{16}\left(\frac{57}{16 b^6}+\frac{591}{112
   b^5}+\frac{8139}{4480 b^4}+\frac{2187}{58240 b^3}-\left(\frac{19}{4b^6}+\frac{39}{4b^5}+\frac{195}{32
   b^4}+\frac{35}{32 b^3}\right) D \right) \nonumber
\end{eqnarray} 
with
\begin{eqnarray}
D = \frac{3}{b} - \frac{3}{b^{4/3}} \bigg[ \frac{1}{6} \log \frac{(1+b^{1/3})^2}{1-b^{1/3}+b^{2/3}}
+ \frac{1}{3^{1/2}} {\rm arctan}\frac{2b^{1/3}-1}{3^{1/3}} + \frac{\pi}{6 \cdot 3^{1/3}}\bigg]
\nonumber
\end{eqnarray}
Incidentally, we mention that the ETF4 SO results of Brack {\it et al.}~\cite{bra85}, based Ref.~\cite{gra80}, are not affected by these mistakes, because they are based on a general expression for the spin-density current, independent of the use of a variational density profile.

The contribution of the spin-orbit at ETF4 order in the case $f=1$ is
\begin{eqnarray}
\frac{E}{S} \bigg|_{SO} &=& \left(\frac{3\pi^2}{2}\right)^{-2/3}\frac{\hbar^2}{2m} \frac{9W_0^2}{16} \left(\frac{m}{\hbar^2}\right)^2 \rho_0^{7/3} \left\{ \frac{9}{728} 
+ \frac{2187}{695552} W_0^2 \left(\frac{m}{\hbar^2}\right)^2 \rho_0^3 
\right\} \alpha^3 \, .
\end{eqnarray}

\subsection{Direct term} \label{subsec:D}
For a Gogny interaction, the contribution of the direct term to the SINM energy density is
\begin{eqnarray}
{\cal{ E}}_D[\rho] & = &  \frac{1}{8} \sum_i \left( 4W_i + 2B_i  - 2 H_i - M_i \right) \rho(z) (\sqrt{\pi} \mu_i)^2 \int_{-\infty}^{\infty} {\rm d}z' \, 
 {\rm e}^{-(z - z')^2 / \mu_i^2} \rho(z')\,,
\end{eqnarray}
while for the homogeneous system it reads
\begin{eqnarray}
{\cal{ E}}_D[\rho_0]& = &  \frac{1}{8} \sum_i \left( 4W_i + 2B_i  - 2 H_i - M_i \right) \rho_0^2 (\sqrt{\pi} \mu_i )^3\,.
\end{eqnarray}
Therefore, the contribution of the direct term to the surface energy reads
\begin{equation}
\frac{E}{S}\bigg|_{D}  = \frac{1}{8} \rho_0^2 
\sum_i ( 4W_i + 2B_i  - 2 H_i - M_i ) (\sqrt{\pi} \mu_i)^2 \int_{-\infty}^{\infty} {\rm d}z \, x(z) 
\left[ \left( \int_{-\infty}^{\infty} {\rm d}z' \, {\rm e}^{-(z - z ')^2 / \mu_i^2} x(z') \right) - \sqrt{\pi} \mu_i \right] .
\end{equation}

\subsection{Exchange term}\label{subsec:E}
The calculation of the exchange term contribution in Eq.~\eqref{eq:Exch} is performed using the various density matrices approaches discussed in Sec.~\ref{Sec:Fock}.  For the SL and CB approximations given in Eqs.~\eqref{eqn:SL} and \eqref{eqn:CB} one can write
\begin{eqnarray}\label{eq:VE}
{\cal E}^{SL,CB}_E[\rho]  
& = &  -\frac{1}{8} \sum_i \left( W_i + 2B_i  - 2 H_i - 4 M_i \right)  \, 
\rho^2 \int d{\bf s} \; \hat \jmath _1(k s)^2 {\rm e}^{-s^2 / \mu_i^2} \nonumber \\
& = &  -\frac{1}{8} \sum_i \left( W_i + 2B_i  - 2 H_i - 4 M_i \right) \, \rho^2  \frac{6 \pi^{3/2}}{k^3 } \, G(k \mu_i)\,.
\end{eqnarray}
The function $G(x)$ is given in Appendix~ \ref{app:A}, and
 $k = k_F(z)$ for the SL case, while $k = \hat k(z)$, as given in Eq.~\eqref{eqn:CB}, for the CB case. The explicit expressions up to ETF4 order of $\hat k$ are also given in the Appendix. Finally, the contribution from the exchange term to the surface energy for these approaches is
\begin{eqnarray}
\label{eqn:SL-CB}
\frac{E}{S}\bigg|_{E}^{SL, CB} &=& - \frac{1}{8}   \sum_i \left( W_i + 2B_i  - 2 H_i - 4 M_i \right) 
\int_{-\infty}^{\infty} {\rm d}z \, x(z) 
\left[ \rho^2 \frac{6 \pi^{3/2}}{k^3 } G(k \mu_i) - \frac{\rho_0}{2 \sqrt{\pi}} G(k_{F0} \mu_i) \right] \nonumber \\
& = &  -\frac{\rho_0}{2\sqrt{\pi}} \sum_i \left( W_i + 2B_i  - 2 H_i - 4 M_i \right)
\int_{-\infty}^{\infty} {\rm d}z \,
\left[ x^2(z) \frac{k_{F0}^3}{k^3 } \, G(k \mu_i)\, -  x(z) \, G(k_{F0} \mu_i) \right]
\end{eqnarray}
The SL expression can be further simplified as
\begin{eqnarray}
\frac{E}{S}\bigg|_{E}^{SL}& = &  -\frac{\rho_0}{2\sqrt{\pi}} \sum_i \left( W_i + 2B_i  - 2 H_i - 4 M_i \right)
\int_{-\infty}^{\infty} {\rm d}z \, x(z)
\left[ G(k \mu_i)\, -  G(k_{F0} \mu_i) \right]
\end{eqnarray}

In the case of the NV approach, the exchange term is written as
\begin{eqnarray}
{\cal E}^{NV}_E[\rho]  & = - & \frac{1}{8} \sum_i \left( W_i + 2B_i  - 2 H_i - 4 M_i \right) \left[ \rho^2 \int d{\bf s} \; \hat \jmath _1(k s)^2 {\rm e}^{-s^2 / \mu_i^2} \right. \nonumber \\
&  & \left. - \frac{1}{3} \rho \left( \frac{1}{4}\nabla^2 \rho - \tau + \frac{3}{5} k^2 \rho \right) \int d{\bf s} \; s^2 \hat \jmath _1(k s) \hat \jmath _3(k_F s) {\rm e}^{-s^2 / \mu_i^2} \right] ,
\label{eq:NV}
\end{eqnarray}
where $k=k_F(z)$. 
Notice that the exchange contribution contains the square of the density matrix, see for example Eq.~\eqref{eq:Exch}. As it has been calculated up to $s^2$, we have omitted the $s^4$ power for consistency. Performing the integrals over $\bf s$ one gets
\begin{eqnarray}\label{eq:NV:Hx}
{\cal E}^{NV}_E[\rho]   & = & - \frac{1}{8} \sum_i \left( W_i + 2B_i  - 2 H_i - 4 M_i \right) \frac{6 \pi^{3/2}}{ k^3 } \left[ \rho^2 \, G(k \mu_i) - \frac{35}{6k^2} \rho \left( \frac{1}{4}\nabla^2 \rho - \tau + \frac{3}{5} k^2 \rho \right) H(k \mu_i)\right].
\end{eqnarray}
The function $H(x)$ is given in the Appendix. 
Finally, the contribution from the exchange term to the surface energy for these approaches can be written as


\begin{equation}\label{eq:NV:Fx}
\frac{E}{S}\bigg|_{E}^{NV} = - \frac{1}{8} \sum_i \left( W_i + 2B_i  - 2 H_i - 4 M_i \right)  \rho_0^2 \int_{-\infty}^{\infty} dz \frac{6 \pi^{3/2} }{k^3} \, 
\bigg\{ x^2(z) G(k \mu_i) - x^2(z)  G(k_{F0} \mu_i) 
 - \frac{35}{6 k^2} F(z)  H(k \mu_i)  \bigg\} ,
\end{equation}
where the expression of function $F(z)$ is given in Appendix~\ref{app:A}  up to ETF4 order.


\section{Results}\label{sec:results}

As mentioned in the Introduction, we take as reference the HF surface energy coefficients calculated in Ref.~\cite{dav25b} and reported in Table~\ref{tab:HF} for the currently used Gogny interactions within the Hartree--Fock approximation. We first note that the inclusion of the zero-range spin-orbit interaction systematically reduces the surface energy coefficient $a_s$ by an average decrease of $1.41 \pm 0.2$ MeV. Furthermore, for reasons that will become clear in the following discussion, we divide the interactions into two distinct groups.
We now discuss the results obtained with the different semiclassical approximations introduced above.

\subsection{Dropping the spin-orbit interaction}
\label{sec:NOSO}
Let us start by ignoring the spin-orbit contribution. The calculated values of $a_s$ are displayed in Table~\ref{table:NOSO}. We differ the analysis of SV results for later on, when discussing non-local effects. 
One can notice that at TF approximation there are no results for the CB approach. As it turns out,  the square of the effective momentum $\hat k$ entering Eq.~\eqref{eqn:CB} is negative, so that the CB approach has no meaning at this level.

\begin{table}[!h]
\begin{center}
\begin{tabular}{l|c|cc|ccc|ccc}
\hline
\hline
\multicolumn{2}{c|}{} & \multicolumn{2}{c|}{TF} & \multicolumn{3}{c|}{ETF2} & \multicolumn{3}{c}{ETF4} \\
\hline
\multicolumn{2}{c|}{}& SL & NV & SL2 & NV2 & CB2  & SL4 & NV4 & CB4 \\
\hline
D1 & $a_s$     & 19.87 &  20.83  &  21.53  & 21.37  &  21.04   & 21.82  
& 21.69 & 21.42 \\
& $\alpha$  & 2.2991 & 2.1486 & 2.0541 & 2.0741 & 2.1335 &  1.9538 & 1.9651 & 1.9968 \\
\hline
D1S  & $a_s$     & 18.47 &  19.33 &   20.05  & 19.93  &  19.68   & 20.31  & 20.22  & 20.03 \\
& $\alpha$   & 2.1840 & 2.0651 & 1.9784 & 1.9914 & 2.0419  & 1.8955 & 1.8998 & 1.9265 \\
\hline
D1N  & $a_s$     & 17.81 & 18.49  &  19.49  &  19.38 &  19.15   & 19.80  & 19.71 & 19.53 \\
& $\alpha$   & 2.3372 & 2.2342 & 2.1008 & 2.1140 & 2.1544   & 1.9969 & 2.0028 & 2.0243 \\
\hline
D1M  & $a_s$     & 18.05  & 18.77  &   19.77  &  19.65 &  19.42   & 20.09  & 20.00 & 19.81 \\ 
& $\alpha$    & 2.3889 & 2.2722 & 2.1312 & 2.1462 & 2.1905   & 2.01960 & 2.02615 & 2.0489 \\
\hline
D1M* & $a_s$    & 18.10 &  18.82 &   19.82  & 19.70  & 19.47   &  20.13  & 20.04 & 19.85 \\
& $\alpha$    & 2.38915 & 2.27251& 2.13177 & 2.1467 & 2.1910   & 2.02029 & 2.02683 & 2.04963 \\
  \hline
D1MK & $a_s$   & 18.11 & 18.83  & 19.85 & 19.73 & 19.49  & 20.18  & 20.09 & 19.90 \\  
& $\alpha$       & 2.4301 & 2.3098 & 2.1641 & 2.1794 & 2.2255  & 2.0461 & 2.0526 & 2.0758 \\
\hline
D3G3 & $a_s$     & 17.48 &  18.28 &  19.32  &  19.26 &  18.95    & 19.67  & 19.65 & 19.42 \\ 
& $\alpha$    & 2.6236 & 2.4241 & 2.2270 & 2.2369 & 2.3498 &   2.0758 & 2.0731 & 2.1234 \\
\hline
D3G3M & $a_s$   & 17.99 &  18.58 &  19.83  &  19.77 &  19.54    & 20.20  & 20.16 & 19.98 \\  
& $\alpha$    & 2.6423 & 2.4876 & 2.2413 & 2.2514 & 2.3225 &  2.0881 & 2.0890 & 2.1226 \\
\hline
\noalign{\vskip 4pt}
     \hline
D250 & $a_s$   & - & - &  - & - & -  & 20.79  & 20.70 &20.49 \\  
& $\alpha$      & - & - & - & - & - &   2.0502 & 2.0543 & 2.0876\\
     \hline
     D260 & $a_s$   & 19.21 & 20.56 & 21.19 & 20.93 & 20.41  &  21.61  & 21.38 & 20.99 \\  
& $\alpha$       & 2.9371 & 2.5056 & 2.3672 &2.4218 & 2.5488  & 2.1653 & 2.1895& 2.2382 \\
\hline
D280 & $a_s$   & - & 20.28   & 20.91 & 20.49 &  19.68 & 21.56  & 21.25 & 20.69 \\  
 &$\alpha$       & - & 3.4126 & 2.9385 & 3.2105 & 4.0660 &  2.4146 & 2.4634 & 2.5214 \\
\hline
D300 & $a_s$   & 24.13 & 25.44 & 26.45 & 26.20    & 25.70  &  27.04 & 26.84 & 26.44 \\  
& $\alpha$      & 3.3368 & 2.9171 & 2.6975 & 2.7463 & 2.8669 &   2.4357& 2.4538 & 2.4945 \\
\hline
D1P & $a_s$     &  17.15 &  18.28 &   19.17   & 18.94 &  18.52  & 19.61 & 19.43   & 19.08 \\
& $\alpha$    & 2.8685 & 2.5925 & 2.4222 & 2.4618 & 2.5469 &  2.2286 & 2.2471 & 2.2830 \\
\hline
D1PK  & $a_s$    & -   & 25.58 & 27.03   & 26.80 & 26.15   &  28.27  & 28.13 & 27.74 \\  
& $\alpha$       & - & 3.8920 & 3.5526 & 3.6012 & 3.8499 &   3.0249 & 3.0273 & 3.7080 \\
\hline
\hline
\end{tabular}
\caption{Surface energy coefficient $a_s$, expressed in MeV, obtained with several Gogny interactions, neglecting the spin--orbit contribution. The results are shown for the Slater, Negele--Vautherin  and Campi--Bouyssy approximations to the density matrix combined with the Thomas--Fermi, ETF2, and ETF4 kinetic-energy densities. For each interaction, the second row gives the value of $\alpha$ (in fm$^{-1}$) that minimizes $a_s(\alpha)$ (see text for details).}
\label{table:NOSO}
\end{center}
\end{table}

For the interactions in the upper half part of the table, we obtain a reasonable agreement at any level of approximation (TF, ETF2 or ETF4) and for all the various approaches presented in the previous section. We also notice that although the values of $\alpha$ differ for each interaction, all of them lies in the relatively small interval between 1.9 and 2.5 fm$^{-1}$.
By construction, the NV approach improves the SL one as it includes the next term in the expansion of the density matrix in powers of $s^2$. Such an improvement is reflected in an increase of $a_s$ ranging from 0.59 to 1.52 MeV at the TF level. The improvement is less important at higher orders : $a_s$ decreases in a range between 0.06 and 0.42 MeV at ETF2 and between 0.02 and 0.31 MeV at ETF4 level. 
As compared to SL and NV approaches, the CB decreases the value of $a_s$ by a similar amount at ETF2 and ETF4 level. 

\begin{figure}[h!]
\begin{center}
\includegraphics[width=0.45\textwidth]{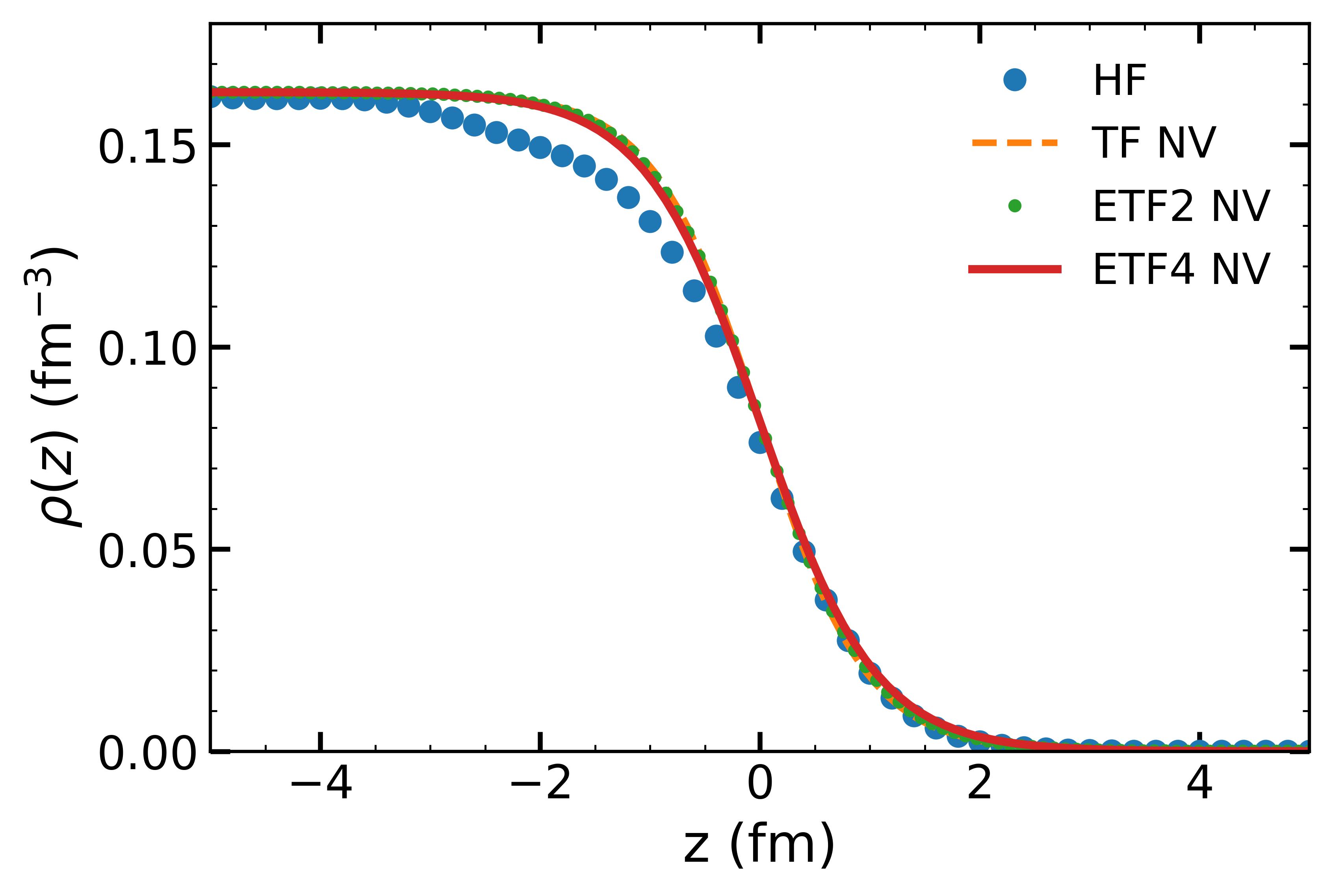}
\includegraphics[width=0.45\textwidth]{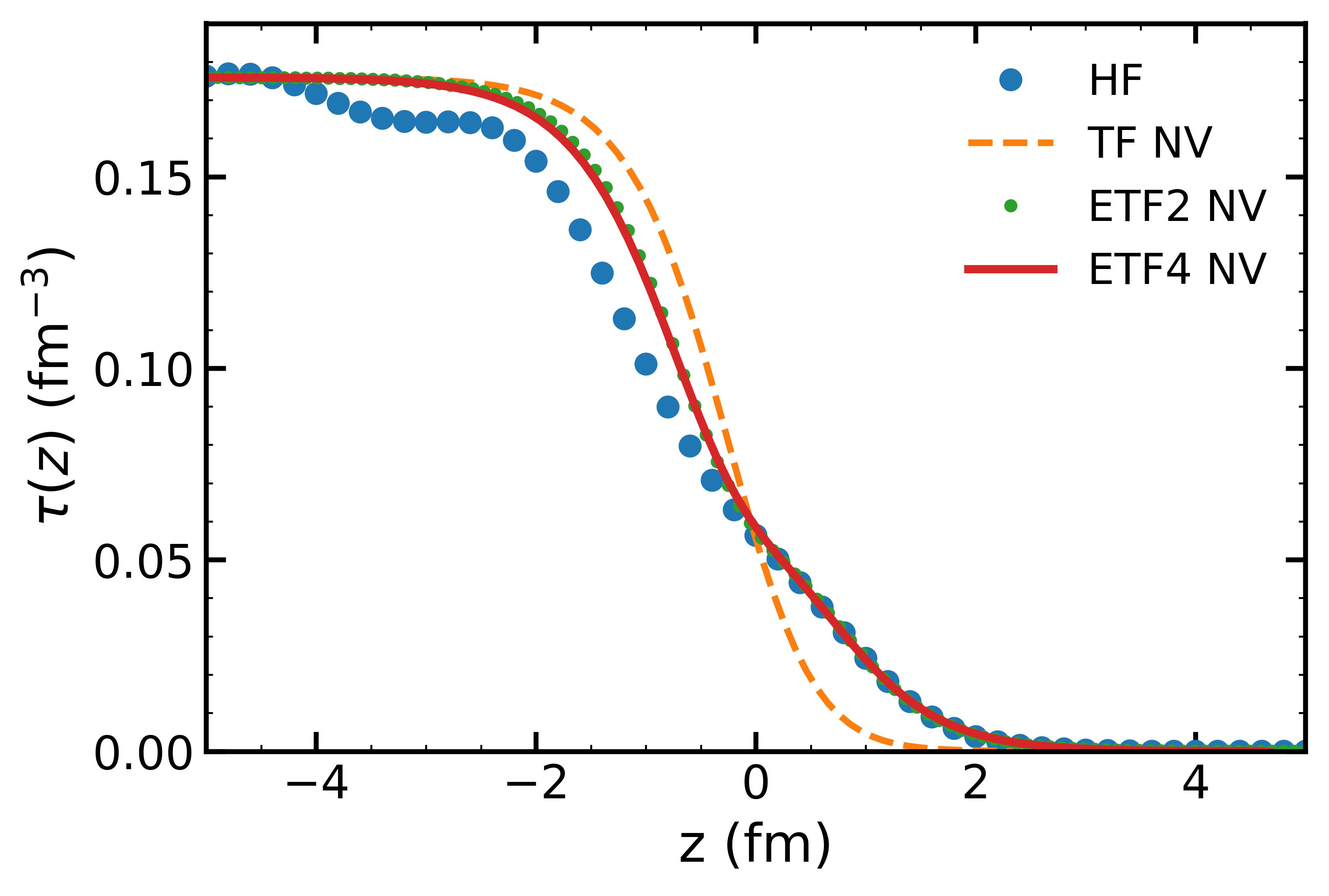}
\caption{Matter density (left panel) and kinetic density (right panel) calculated using the D1S interaction without spin-orbit term for the full HF case \cite{dav25b} and various order of semiclassical approximations using the Negele and Vautherin prescription. See text for details. }
\label{fig:rho:tau}
\end{center}
\end{figure}

In Fig.~\ref{fig:rho:tau}, we compare the HF matter and kinetic-energy densities with their semiclassical counterparts obtained within the Negele--Vautherin (NV) approach at the TF, ETF2, and ETF4 levels, using the D1S interaction. We observe that the TF approximation already reproduces the main features of both densities, while the inclusion of the second-order ETF corrections yields a substantial improvement of the kinetic-energy density, bringing it into good agreement with the HF result. The additional fourth-order corrections lead only to marginal changes.
On the other hand, neither the ETF2 nor the ETF4 approximation is able to reproduce the oscillations that develop in the surface region ($z<0$) : these Friedel oscillations are purely quantal in origin and are therefore absent from the present semiclassical description. Moreover, the adopted variational density profile is essentially flat in this region, which further prevents the appearance of such density fluctuations.

It is worth noticing that, in Table~\ref{table:NOSO},  the values of $\alpha$ are very similar for the first group of interactions within each level of approximation to the kinetic energy. Actually, the average values are  $\overline{\alpha}_{TF}=2.36 \pm0.15$ fm$^{-1}$, $\overline{\alpha}_{ETF2}=2.17 \pm 0.09$ fm$^{-1}$, and $\overline{\alpha}_{ETF4}=2.03 \pm 0.06$ fm$^{-1}$. In other terms, the surface width of the variational density profile increases as 1.86, 2.03 and 2.16 fm$^{-1}$, respectively.

\begin{figure}[h!]
\begin{center}
\includegraphics[width=0.45\textwidth]{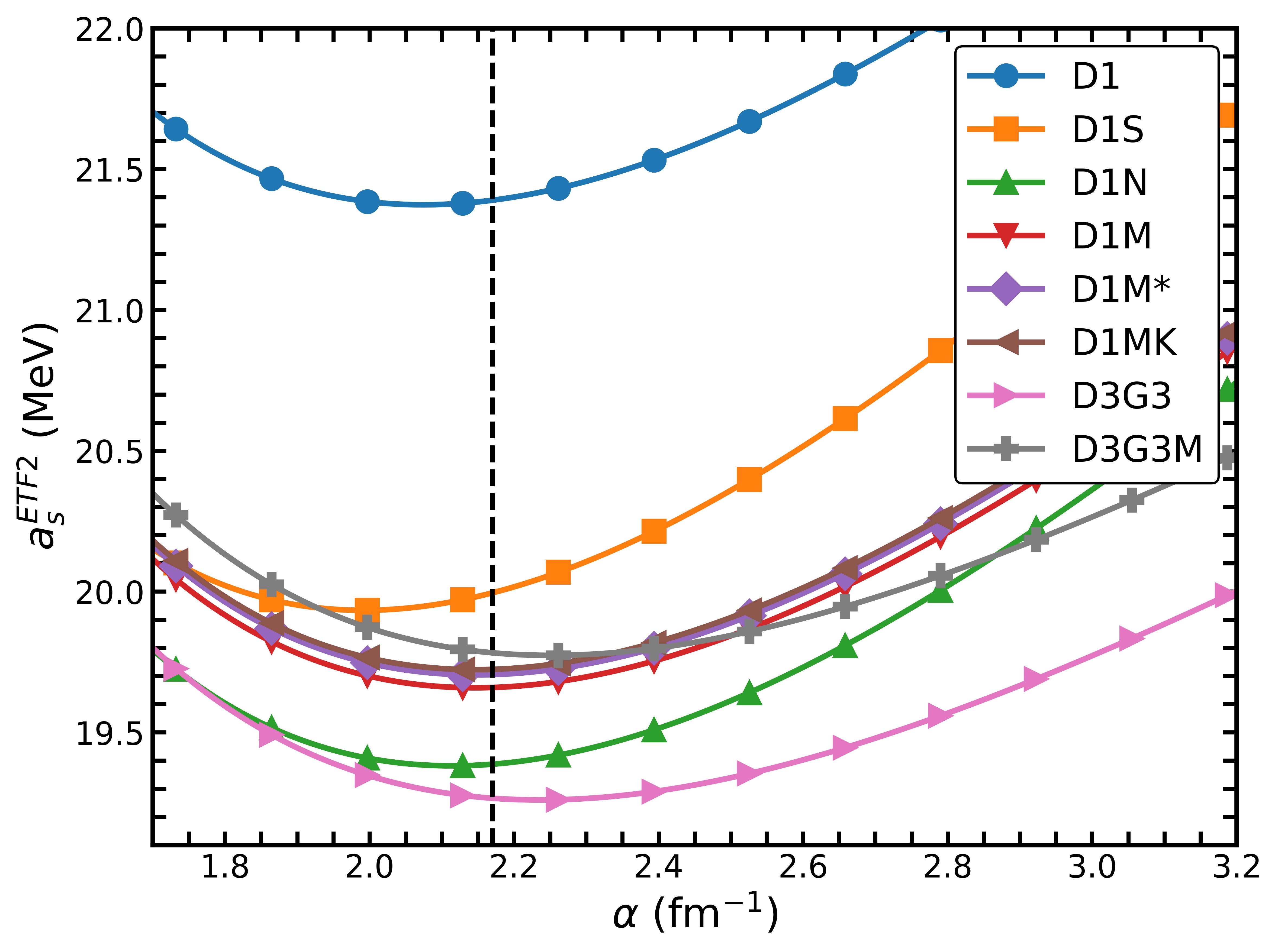}
\caption{Surface energy coefficient $a_s^{\mathrm{ETF2}}(\alpha)$ (in MeV) as a function of the diffuseness parameter $\alpha$ (in fm$^{-1}$), calculated within the NV approach at the ETF2 level for the first group of Gogny interactions, omitting the spin-orbit term. See the text for details.}
\label{fig:alpha}
\end{center}
\end{figure}

In Fig.~\ref{fig:alpha}, we plot the surface energy coefficient $a_s(\alpha)$ as a function of $\alpha$, calculated within NV2 approach for the first group of  interactions. As a matter of fact, the minimum of $a_s$ lies in a relatively flat region. We have found that this behaviour is similar for all interactions and all approaches. The vertical line in the figure corresponds to $\alpha=2.17$ fm$^{-1}$, which is the average value for all interactions and approaches at the ETF2 level. One can check that calculating $a_s$ with this value of $\alpha$ gives a result which is very close to the true minimum energy. However, as can be appreciated in Table~\ref{table:NOSO}, such an average value depends on the level of approximation for the kinetic energy.

Let's consider now the interactions belonging to the second group. No values of $a_s$ are given in Table~\ref{table:NOSO} for some interactions, for which the curve $a_s(\alpha)$ decreases monotonically as $\alpha$ increases, with either no minimum or reaching a plateau at a very high value of $a_s$ and very large values of $\alpha$. The results are not significant and not reported here. 
The remaining interactions lead to values of $\alpha$ which are larger than those of the first group, lying in the interval 2.1 and 3.3 fm$^{-1}$. Notice that interaction D1PK produces too high values both for $a_s$ and $\alpha$.

\begin{figure}[h!]
\begin{center}
\includegraphics[width=1\textwidth]{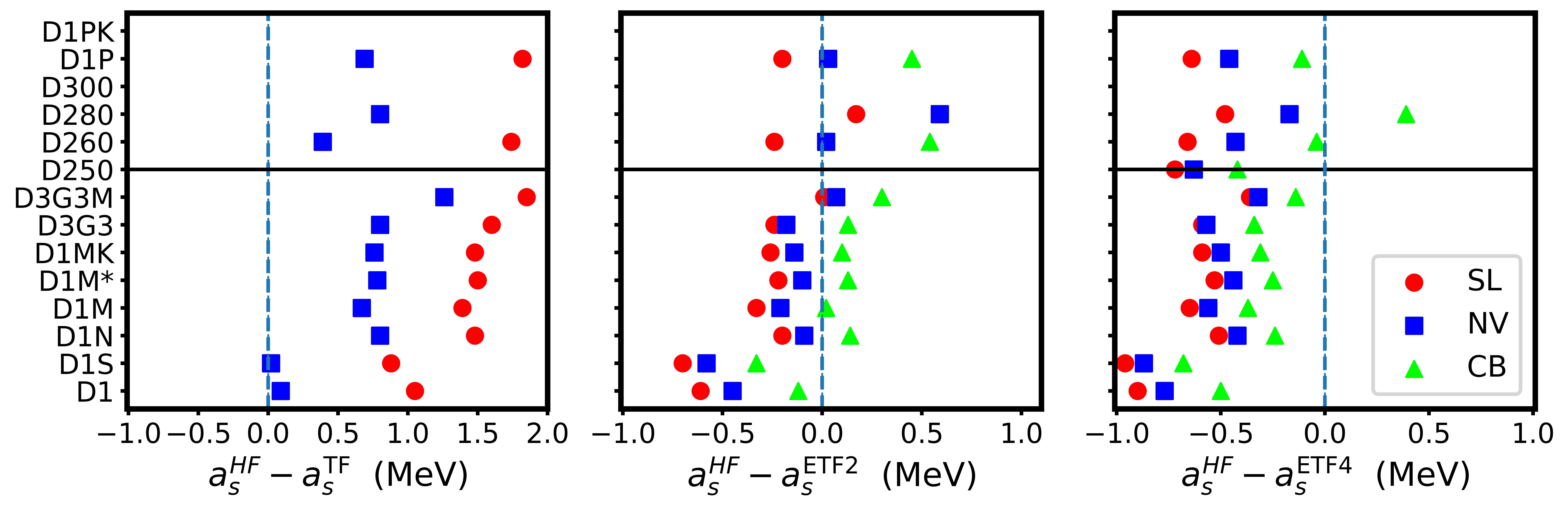}
\caption{Differences between HF and TF (left panel), ETF2 (middle panel) and ETF4 (right panel) for $a_s$ when spin-orbit term is not included in the calculation. The solid horizontal line separates the two families of interactions as discussed in the text. }
\label{fig:NOSO}
\end{center}
\end{figure}

In Fig.~\ref{fig:NOSO}, we plot the difference between the HF surface energy coefficient and the values obtained at  TF, ETF2 and ETF4 levels, using SL, NV and CB approaches, when the spin-orbit interaction is dropped. One notices that ETF2 calculations results in differences of less than 0.5 MeV. As compared to SL and CB approaches, the $a_s$ values given by CB results are closer to HF ones. 

\subsection{Spin--orbit effects}
\label{sec:SO}

\begin{figure}[h!]
\begin{center}
\includegraphics[width=0.45\textwidth]{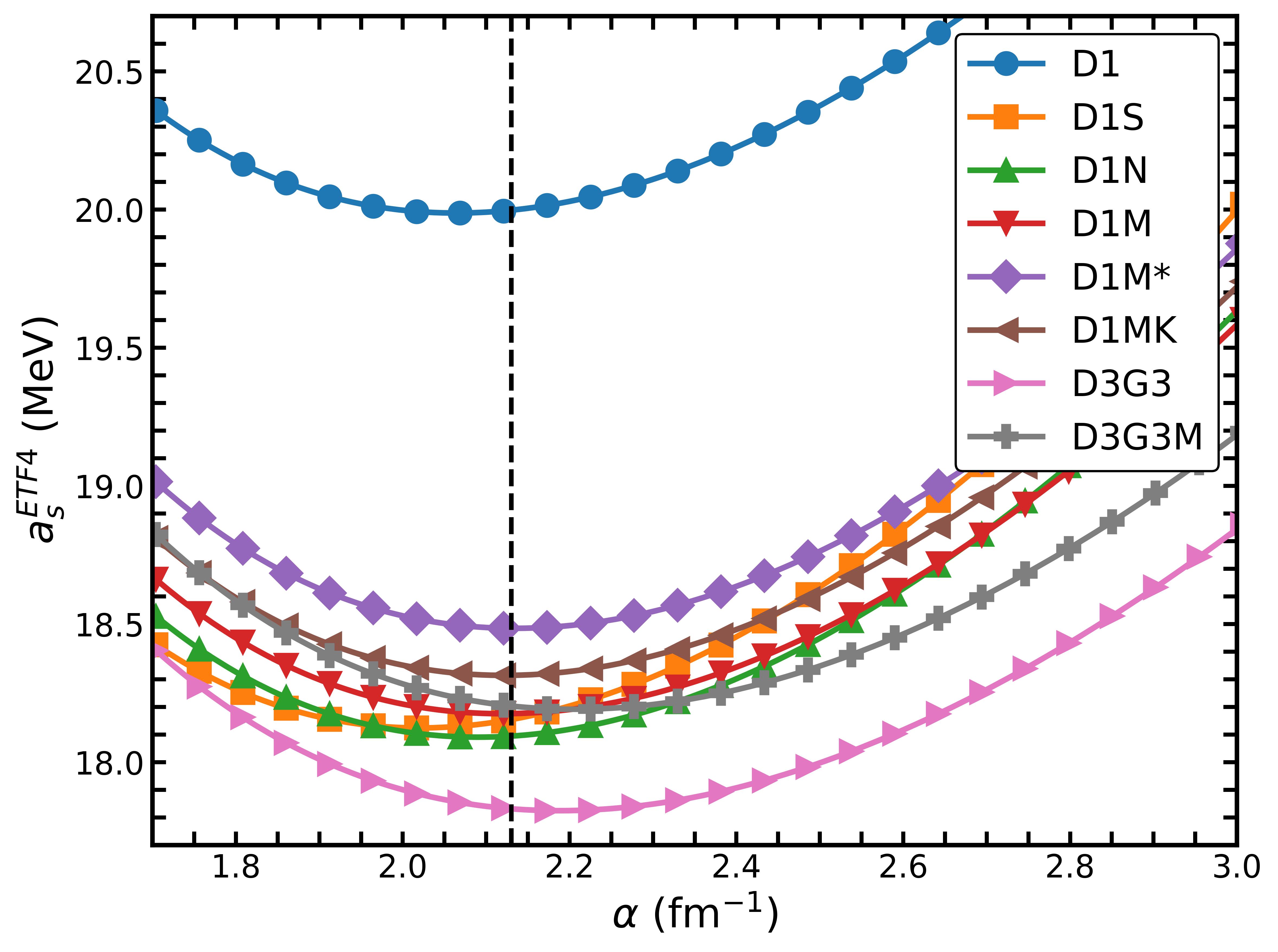}
\includegraphics[width=0.45\textwidth]{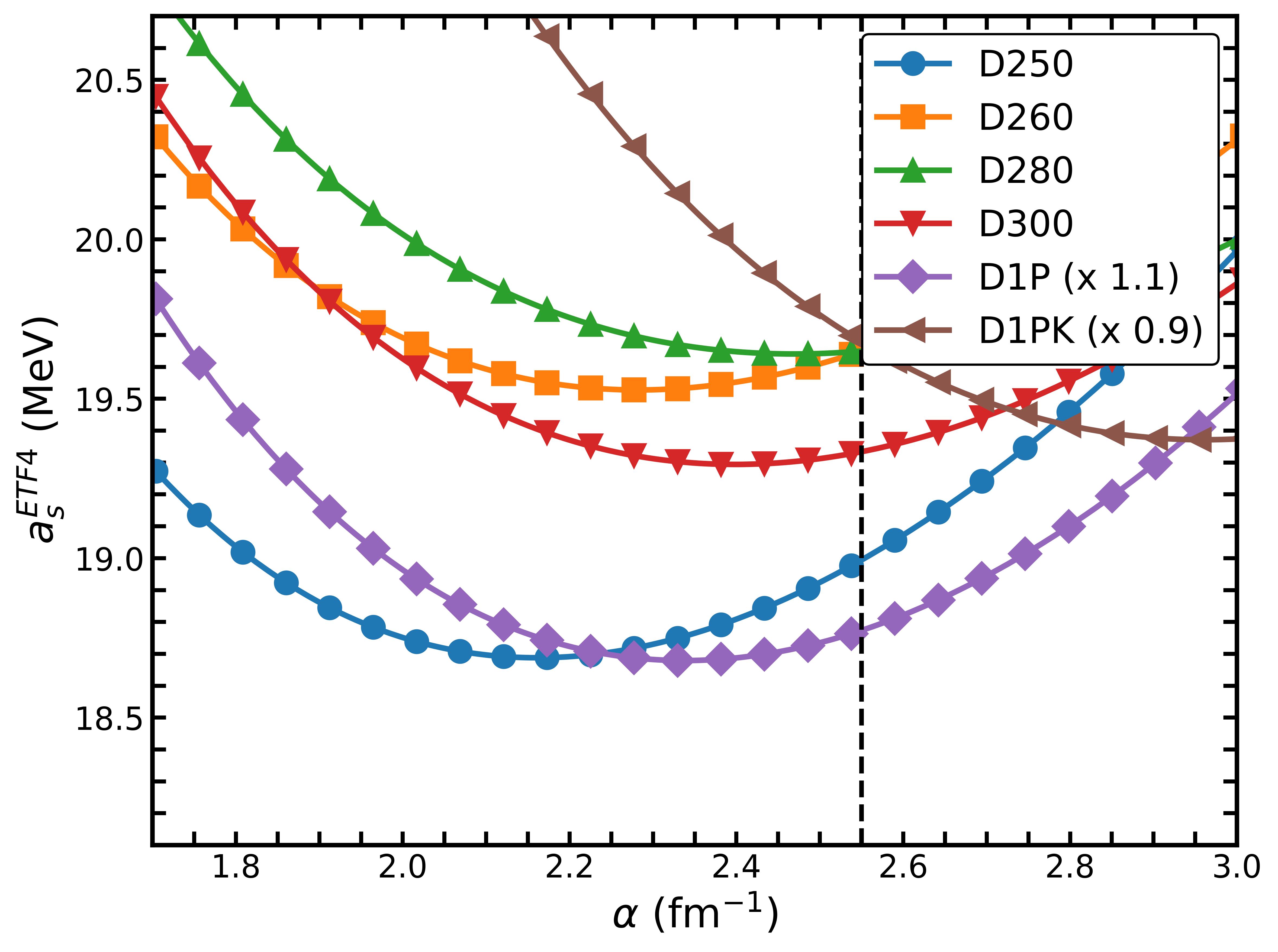}
\caption{Same as Fig.\ref{fig:alpha}, but for the case including spin orbit. Left panel: interactions on the first group; right panel: interaction of the second group. The vertical dashed line correspond to the average value of $\alpha$.}
\label{fig:alpha:SO}
\end{center}
\end{figure}

We now discuss the values of $a_s$ obtained when the spin--orbit contribution is included in the interaction. The results obtained by minimizing $a_s(\alpha)$ are presented in Table~\ref{table:SO}.
Notice that interactions of the second group at the ETF2 level fail to exhibit a variational minimum, except D1P within SL2 approach. By comparing Tables~\ref{table:NOSO} and~\ref{table:SO}, one can see that the spin--orbit term reduces the value of the surface energy coefficient $a_s$. At ETF2 level, the reduction lies between 2-3 MeV, roughly 1 MeV stronger than the reduction observed at HF level when this term is included.
At ETF4 level, all interactions considered here lead to a converged result. However, for the ones where an ETF2 result exist, we notice that, as in the previous case, the change from ETF2 to ETF4 is small and roughly of the order of 0.5 MeV. This effect is somehow larger than what observed in the case with no spin--orbit, thus showing that the ETF4 is relevant to have a correct insight about the spin-orbit effect.
Similarly to the case with no spin--orbit, the values of $\alpha$ are very close for the first group of interactions for a given level of approximation of the kinetic energy.  
In Fig.~\ref{fig:alpha:SO}, we show the evolution of $a^{ETF4}_s$ in function of $\alpha$ for both groups of interactions. We see that  the minimum of $a_s$ as a function of $\alpha$ lies in a relatively flat region around the averaged minimum values   $\alpha= 2.13 $ fm$^{-1}$ for ETF4. This value is reported as a vertical dashed line on the figure.
The interactions belonging to the second group on the contrary exhibit a different behavior : the average value of $\alpha= 2.55 $ fm$^{-1}$ for ETF4 is also reported, but we see that in this case the variance is much higher and this value makes little sense now. We also had to rescale the value of D1P and D1PK in order to fit them in the figure.

\begin{table}[!h]
\begin{center}
\begin{tabular}{|l|c|ccc|ccc|}
\hline
 \multicolumn{2}{|c|}{}&  \multicolumn{3}{|c|}{ETF2} & \multicolumn{3}{|c|}{ETF4} \\
\hline
 \multicolumn{2}{|c|}{}&  SL2 & NV2 & CB2  &  SL4 & NV4 & CB4 \\
\hline
D1   &$a_s$   &   19.07  & 18.89 & 18.45   & 19.99 & 19.85  & 19.56  \\
 &$\alpha$& 2.4506 & 2.4090 & 2.6432  & 2.0641 & 2.0734 & 2.1064 \\
\hline
D1S  &$a_s$   &   17.14  & 17.00& 16.64   &18.12 & 18.03  & 17.82   \\
&$\alpha$ & 2.4066 & 2.4331 & 2.5768  & 2.0200 & 2.0237 & 2.0530 \\
\hline
D1N &$a_s$   &  17.18  & 17.05 & 16.77  & 18.09& 18.00  & 17.81  \\
&$\alpha$ & 2.4455 & 2.4684 & 2.5507  & 2.0906 & 2.0961 & 2.1182 \\
\hline
D1M &$a_s$    &  17.23  & 17.13  & 16.82  & 18.18 & 18.08  & 17.88  \\
&$\alpha$ & 2.5691 & 2.5647 & 2.6677  & 2.1189 & 2.1250 & 2.1483 \\
\hline
D1M*&$a_s$  &    17.30 & 17.16 & 16.85  & 18.44 & 18.39  & 18.18  \\
&$\alpha$ & 2.5384 & 2.5667 & 2.6699  & 2.1347& 2.1409 & 2.1646 \\
 \hline
 D1MK&$a_s$ &  17.36 & 17.22 & 16.90   &  18.26 & 18.22  & 18.02 \\
&$\alpha$ & 2.5740 & 2.6030 & 2.7102 &   2.1219 & 2.1280 & 2.1513 \\
\hline
D3G3&$a_s$   &  16.61 &  16.53  & -           &  17.82 & 17.80  & 17.55 \\
&$\alpha$ & 2.9416 & 2.9790 & - &  2.1882 & 2.1843 & 2.2366 \\
\hline
D3G3M&$a_s$ &  16.84 & 16.76 & - &  18.19 & 18.15  & 17.96 \\
&$\alpha$ & 3.1314 & 3.1938 & - &   2.2059 & 2.2059 & 2.2404 \\
\hline
\hline
D250&$a_s$ &  - &     -      & -         &  18.69 & 18.60  & 18.37 \\
&$\alpha$ & - & - & - &   2.1611 & 2.1643 & 2.1985 \\
\hline
D260&$a_s$ &  - &     -      & -         &  19.53 & 19.29  & 18.85 \\
&$\alpha$ & - & - & - &   2.2821 & 2.3041 & 2.3479 \\
\hline
D280&$a_s$ & - &     -      & -         &  19.64 &  19.32 & 18.75\\
&$\alpha$ & - & - & - &  2.4695 & 2.5537 & 2.5690  \\ 
\hline
 D300&$a_s$ &  - &     -      & -         &  19.29 & 19.12  & 18.78 \\
&$\alpha$ & - & - & - &  2.3986 & 2.4135 & 2.4462 \\
\hline
D1P&$a_s$ & 14.89 &     -      & -         &  16.98 & 16.78  & 16.42 \\
&$\alpha$ & 4.0526 & - & - &  2.3381 & 2.3544 & 2.3858 \\ 
\hline
D1PK&$a_s$ &  - &     -      & -         &  25.83 & 25.69  & 25.31 \\
&$\alpha$ & - & - & - &  2.9609 & 2.9632 & 2.9987 \\
\hline
\hline
\end{tabular}
\caption{Surface energy coefficient $a_s$ (in MeV) calculated with several Gogny interactions including the spin--orbit contribution. For each interaction, the value of the variational parameter $\alpha$ (in fm$^{-1}$) corresponding to the minimum of $a_s(\alpha)$ is given in the row below.
}
\label{table:SO}
\end{center}
\end{table}

\begin{figure}[h!]
\begin{center}
\includegraphics[width=1\textwidth]{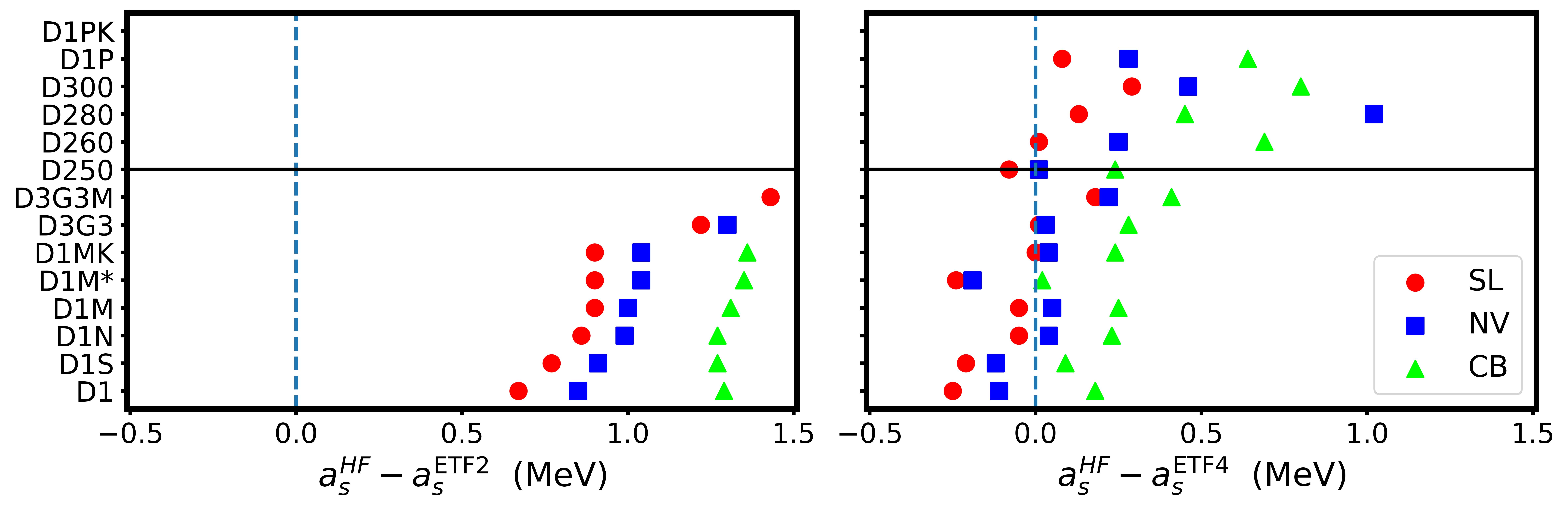}
\caption{Same as Fig.~\ref{fig:NOSO} when the spin-orbit term is included.}
\label{fig:SO}
\end{center}
\end{figure}

In Fig.~\ref{fig:SO}, we show the differences with respect to the HF surface energy coefficient of the values obtained at the ETF2 (left panel) and ETF4 (right panel) levels, using SL, NV and CB approaches.
At ETF2 the results (when they exist) are very much vertically aligned, showing almost a constant shift of $\approx$ 0.7 MeV for SL, $\approx$ 1 MeV for NV and $\approx$ 1.3 MeV for CB. In all cases the semiclassical approximation underestimates the HF value. 
Moving to ETF4 and leaving aside the interaction of the second group, we clearly notice that the SL and NV results are very close to each other, while the CB values are typically shifted to the right by 100 to 200 keV compared to the NV systematic.
We notice that the the semiclassical results, independently on the approximation we used, are capable to reproduce the HF within  $\approx$0.5 MeV accuracy.
This result is in agreement with previous findings based on Skyrme interaction done in Ref.~\cite{jod16} using a larger data-set.
The behaviour of D3G3M is somehow in between the two groups, the previous conclusion still apply apart from a more pronounced shift of the semiclassical result.
Considering now the second group interactions, we notice that the semiclassical approximation presents a large variability depending on the adopted approximation, but realistically they still reproduce HF results although with an error which can be as large as 1.2 MeV, depending on the approach and the interaction.

\begin{figure}[h!]
\begin{center}
\includegraphics[width=1\textwidth]{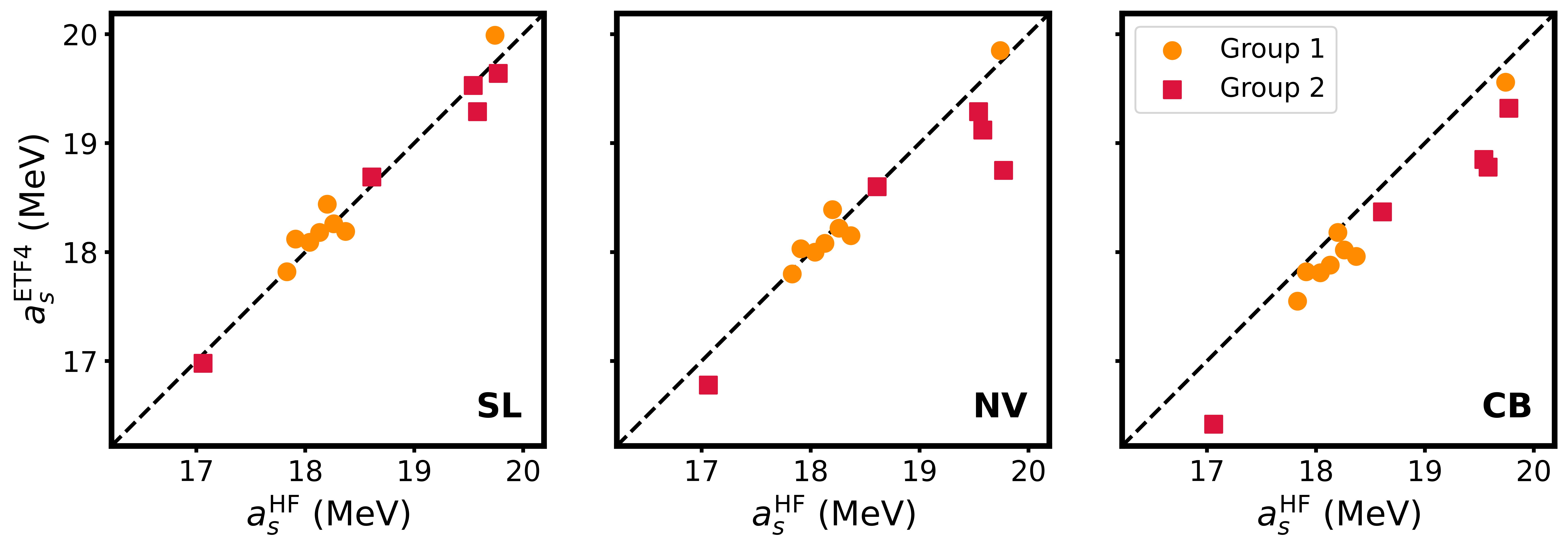}
\caption{Scatter plot of the semiclassical ETF4 surface energies against the HF ones. The interactions are grouped in two families as discussed in the text. The dashed lines represent the diagonal and help guiding the eye.}
\label{fig:SO:HF}
\end{center}
\end{figure}

In Fig.~\ref{fig:SO:HF}, we show a scatter plot of the HF surface energy coefficients calculated including spin-orbit and the corresponding semiclassical ones calculated at ETF4 and three different approximations of the density matrix.
In this figure, we just separate the two groups of interactions. We observe that for the interactions of the first group the points lie essentially on the diagonal for all three approximations independently on the value of $a_s$. The semiclassical approximation does not depend on the strength of the surface coefficient. For the interaction of the second group we notice that the semiclassical approach underestimates systematically the quantal result apart from the SL approximation where the points are way closer to the diagonal.
In all these figures, the value of D1PK has been neglected since is completely out of scale as shown in Tab.\ref{table:SO}.

In conclusion, the smaller differences between semiclassical and HF results are obtained when the spin--orbit is included and ETF4 approaches are used. Another way to visualize this agreement is shown in Fig.~\ref{fig:SO:HF}, where the three semiclassical ETF4 surface energies are plotted against the HF ones. Each panel corresponds to one of the semiclassical approximations to the density matrix. Each point represents one of the interactions employed here, excluding D1PK. The dashed line is the ideal case where both semiclassical and HF results coincide. One can see that SL approach produces the better agreement with HF results, even for the second group of interactions. For these interactions, NV and CB approaches give a worst agreement.  

\subsection{Discussion of Friedel oscillations}

\begin{figure}[h!]
\begin{center}
\includegraphics[width=0.45\textwidth]{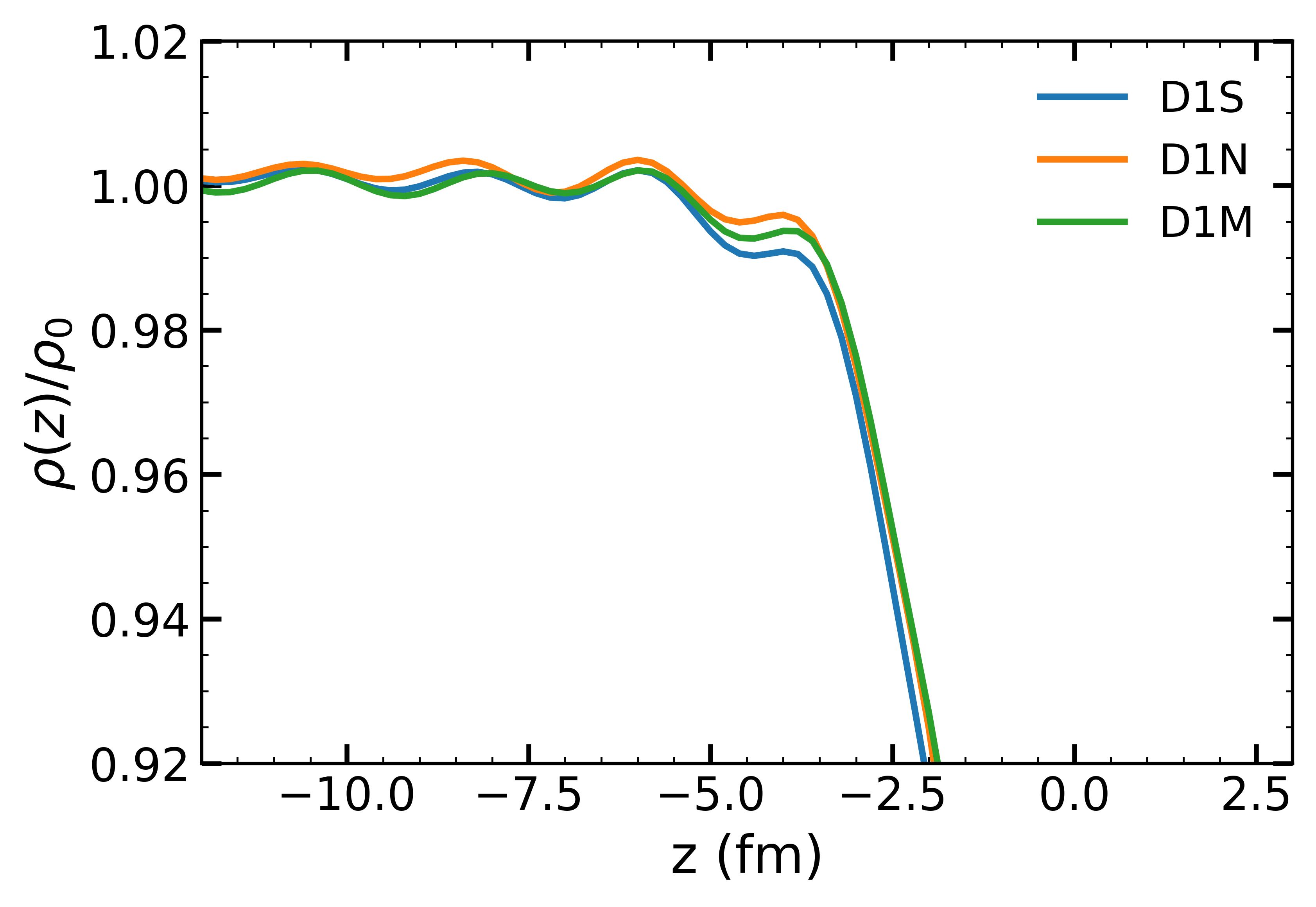}
\includegraphics[width=0.45\textwidth]{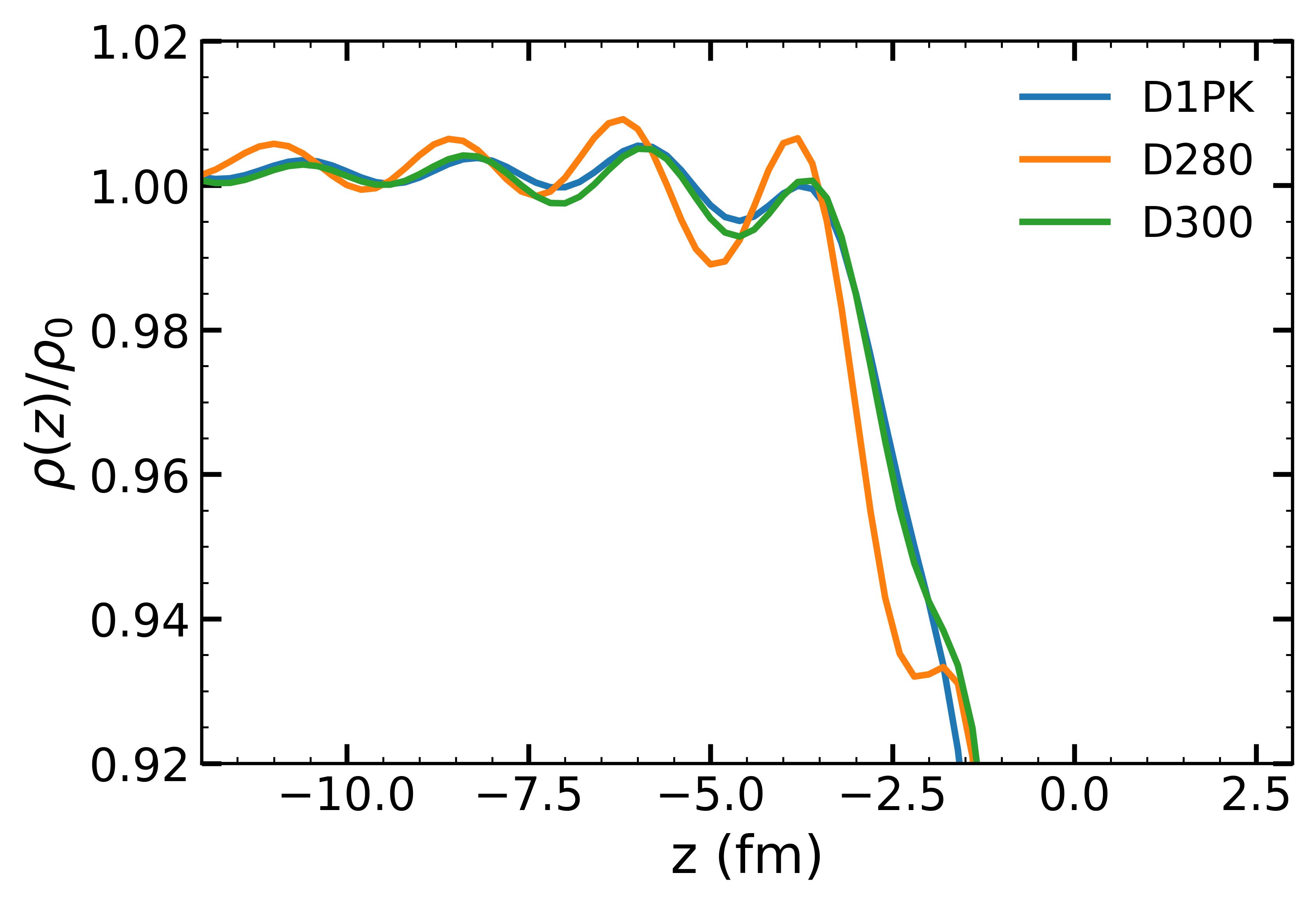}
\caption{Rescaled  density profiles for interactions belonging to the first group, namely D1 D1S and D1N (left panel) and to the second group, namely D1PK, D280 and D300 (right panel) obtained using complete HF  calculations of Ref.\cite{dav25b}. All calculations have been done with no spin orbit term. 
}
\label{fig:rho:compare}
\end{center}
\end{figure}

\begin{figure}[h!]
\begin{center}
\includegraphics[width=0.45\textwidth]{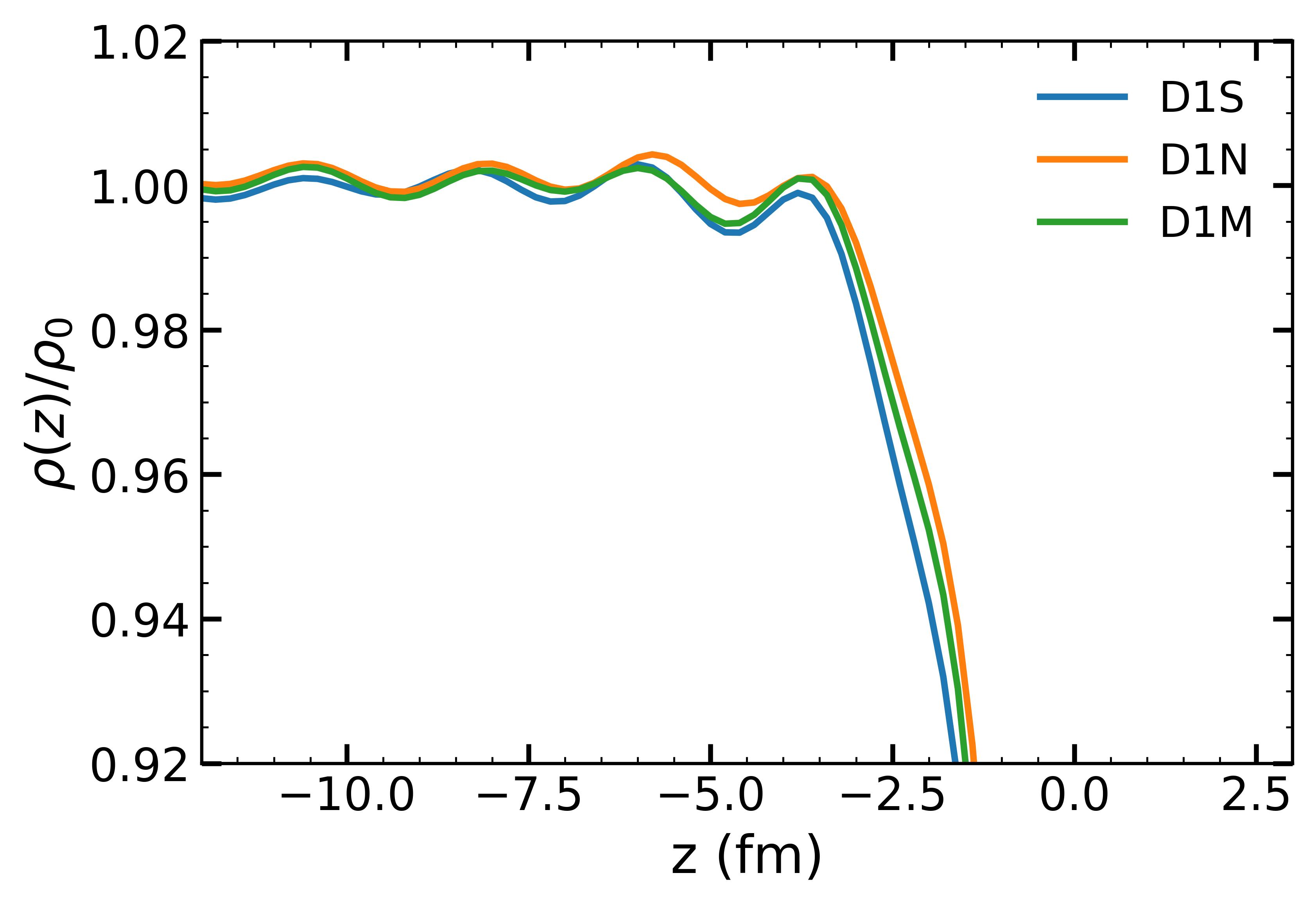}
\includegraphics[width=0.45\textwidth]{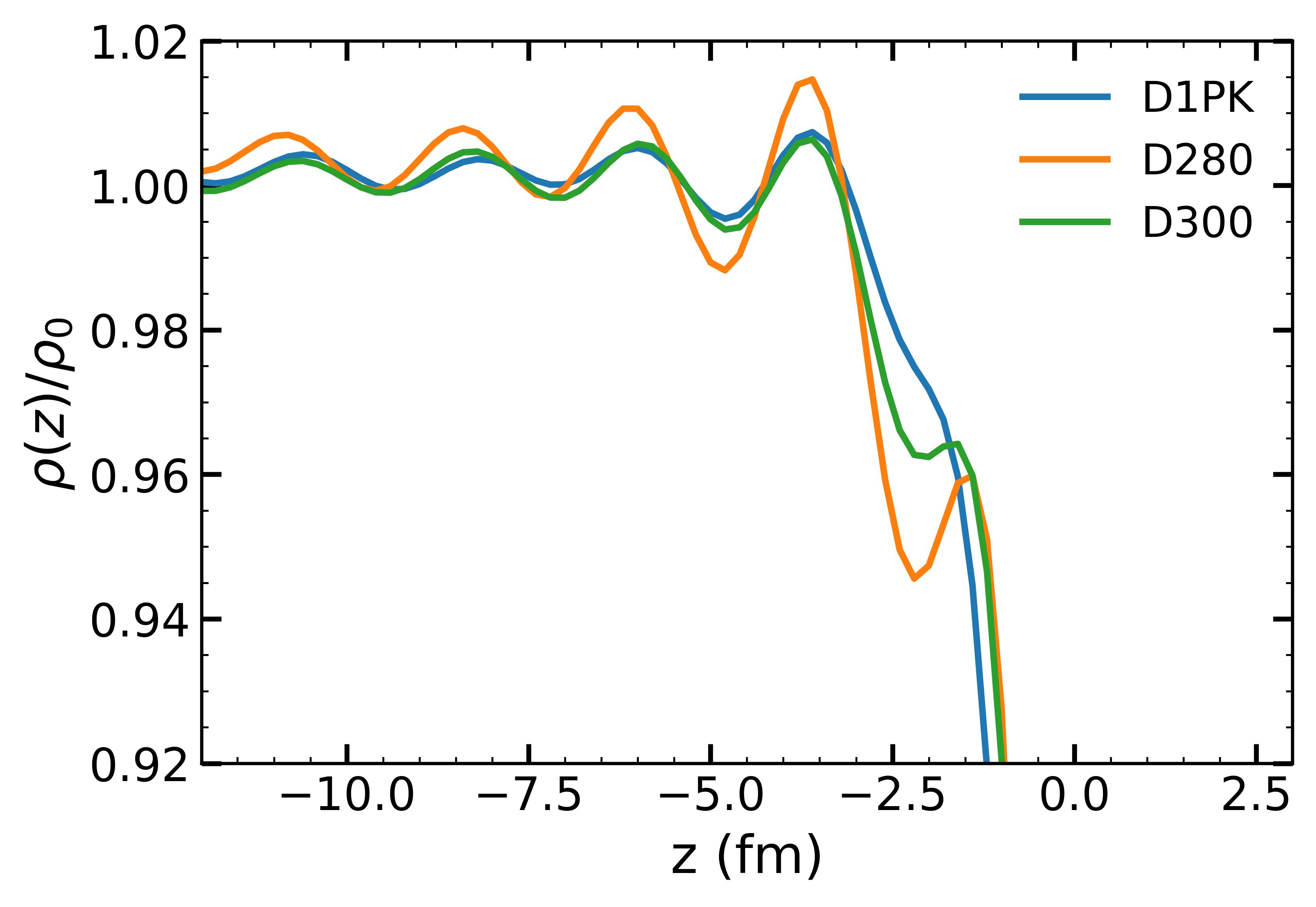}
\caption{Same as Fig.\ref{fig:rho:compare}, but including explicitly spin-orbit term. 
}
\label{fig:rho:compareSO}
\end{center}
\end{figure}

We have seen that the largest discrepancies are observed for the second group of interactions. To this respect, it is worth recalling that, as mentioned in the Introduction, these interactions do not describe finite nuclei at the same level of accuracy as those of the first group. 
To better understand the different behavior of both groups, we display in  Figs.~\ref{fig:rho:compare} and~\ref{fig:rho:compareSO} densities in the surface region obtained from a full Hartree--Fock calculation without and with the spin--orbit contribution, respectively. The densities are rescaled to their saturated values in INM, and the figure is enlarged to highlight the presence of surface oscillations.
The left panels display three representative interactions from the first group, whereas the right panels show three representative interactions from the second group. The most striking feature is the enhancement of the Friedel oscillations for the interactions belonging to the second group.

We recall that these oscillations arise from an interference between forward and backward wavefunctions at the surface, creating a standing wave pattern in the density profile which dies off inside the matter due to phase cancellation. Their typical length scale $L_F$ is of order $\pi/k_F \simeq 2.3$ fm at saturation density. One expects that these oscillations are more important for stiffer surface density. 
It is therefore clear that the simple density profile defined in Eq.~\eqref{density:prof:SINM} is unable to reproduce these additional oscillations, and it is not surprising that the accuracy of our semiclassical approach deteriorates for these interactions. Interestingly, the values of $\alpha$ giving the minimum surface energy are larger for interactions of the second group, which means a smaller surface width and thus a stiffer density.

In order to be able to better characterise each of these two groups, we summarise in Tab.V, 
some features associated with these interactions, as given by a Hartree--Fock calculation. In particular we provide the explicit contributions to the energy per particle of SNM at saturation density $E/A$ :
the kinetic energy per particle ($E_T/A$), the direct and exchange contribution ($E_D/A$,$E_E/A$) and the density dependent term ($E_{DD}/A$), these last three quantities being divided by $E_T/A$, which is practically the same for all interactions.  
Since the semiclassical approximation is done over the Fock term, we may expect to have better results for the interactions that receive a small contribution from this exchange term. This is exactly what is observed for all the interactions of the second group apart from the D3 family where the Fock term being typically one order of magnitude larger that for the other interactions. 
The Fock term is also large for the interaction of the second group. Notice the different density-dependent contributions $E_{DD}/E_T$ for each group: larger than 2 for the first one, lesser for the second one.  
In the last column of Tab.V, 
we also report the nuclear incompressibility $K$~\cite{bla95}. Interestingly, the interactions for which the semiclassical approach provides the best agreement with the HF surface energy coefficients are characterized by relatively low incompressibilities, in the range $K \simeq 200$--220 MeV. In contrast, the interactions belonging to the second group were specifically designed to explore significantly larger values of $K$.

With the present data set, however, it is not possible to determine whether this correlation is merely accidental or whether the incompressibility is indeed responsible for the pronounced density oscillations. Addressing this question would require the construction of a dedicated family of Gogny interactions in which the incompressibility could be varied independently while carefully controlling the relative contribution of the Fock term. Such an investigation lies beyond the scope of the present work and is left for future studies.

\begin{table}[!h]
\begin{center}
\caption{Characteristics of the interactions in SNM based on HF results. See text for details.}
\begin{tabular}{|l|c|cc|c|c|c|}
\hline
& \multicolumn{5}{|c|}{Contributions to $E/A$ (MeV)} & \\
\hline
 & $E_T/A$ (MeV) & $E_D/E_T$ & $E_E/E_T$ & $(E_D+E_E)/E_T$ & $E_{DD}/E_T$ & $K$ (MeV) \\
 \hline
 D1S   & 22.39 & -3.69 & -0.01 & -3.70 & 2.07 & 203 \\
 D1    & 22.70 & -2.40 & -1.36 & -3.76 & 2.04 & 230 \\
 D1M   & 22.33 & -4.01 & -0.04 & -4.05 & 2.36 & 225 \\
 D1M*  & 22.50 & -4.02 & -0.04 & -4.06 & 2.36 & 225 \\
 D1MK  & 22.41 & -4.03 & -0.35 & -4.38 & 2.34 & 225 \\ 
 D1N   & 22.22 & -3.32 & -0.70 & -4.02 & 2.38 & 226 \\
 D3G3  & 22.55 & +3.09 & -6.90 & -3.81 & 2.10 & 227 \\
 D3G3M & 22.54 & +4.09 & -8.21 & -4.12 & 2.40 & 240 \\
 \hline
D250   & 21.92 & -2.66 & -0.13 & -2.79 & 1.07 & 250 \\

D260   & 22.03 & -0.28 & -3.36 & -3.64 & 1.92 & 260 \\
D280   & 21.32 & +1.67 & -5.28 & -4.01 & 1.86 & 285 \\
D300   & 21.75 & -0.49 & -2.39 & -2.88 & 1.13 & 299 \\
D1P    & 22.91 & -1.68 & -1.69 & -3.36 & 1.71 & 254 \\
D1PK   & 22.41 & -2.27 & -1.15 & -3.42 & 1.70 & 260 \\
\hline
\end{tabular}
\end{center}
\label{tab:SNM_contrib}
\end{table}

\section{Non-local effects}\label{sec:NL}
In the previous sections, we have ignored non-local effects on the kinetic density since we have decided to act at various levels of approximation only on the exchange term. In the case of the zero-range Skyrme interaction, Grammaticos and Voros~\cite{gra79} have shown that non-local effects have a small impact on the surface energy. For instance, for SIII interaction they found that the difference in using Eqs.~\eqref{eq:tau4} and~\eqref{eq:tau2*} is around 0.08 MeV at $\hbar^2$ order. To asses the importance of non-local effects in the case of finite-range interactions, we consider another semiclassical approach to the density matrix, developed by Soubbotin and Vi\~nas (SV) in Ref.~\cite{sou00}. This approximation is quite general and can be applied to any finite-range force to describe finite-nuclei at ETF2 level. Besides, shell effects, absent in a semiclassical approach, have been recovered in a very efficient way in the 
spirit of the Kohn-Sham scheme by replacing in the ETF energy density functional the semiclassical particle, kinetic energy and spin densities by the HF ones. This method was introduced formally in Ref.~\cite{sou03} through a quasilocal reduction 
of the non-local density functional theory and generalized to take pairing correlations in Ref.~\cite{kre06}. It is found that this approach reproduce extremely
 well the full HF or HFB calculations, as it can be seen in Refs.~\cite{sou03,kre06,beh16}.  In the following, we shall present the relevant results in order to calculate the SINM surface energy.

\subsection{ETF approximation with finite-range forces}
The ETF approximation to the HF energy for non-local potentials consists of replacing the quantal HF density matrix by its semiclassical
counterpart, which is obtained from the Wigner-Kirkwood expansion of the distribution function (Wigner transform of the density matrix~\cite{ring:book}) 
given by Eq.(A.9) of Ref.~\cite{sou00}.
This ETF density matrix for each kind of nucleons reads:
\begin{widetext}
\begin{eqnarray}
{\tilde \rho}_{SV}\left({\bf R},s\right) &=& \frac{3j_1\left(k_F s\right)}{k_F s}\rho
+ \frac{s^2}{216}\Bigg[\left\{\left[\left(9 - 2k_F\frac{f_k}{f}
- 2k_F^2\frac{f_{kk}}{f} +  k_F^2\frac{f_k^2}{f}\right)\frac{j_1\left(k_F s\right)}{k_F s} -
4 j_0\left(k_F s\right)\right]
\frac{\left({\bf \nabla}\rho\right)^2}{\rho}  \right. \nonumber \\
&& \left. - \left[\left(18 +  6k_F\frac{f_k}{f}\right) \frac{j_1(k_F s)}{k_F s} - 3j_0(k_F s)\right]\Delta \rho \right. \nonumber \\
&& \left.
- \left[18\rho \frac{\Delta f}{f} + \left(18 - 6k_F \frac{f_k}{f}\right)
\frac{{\bf \nabla}\rho \cdot {\bf \nabla} f}{f}
+ 12 k_F \frac{{\bf \nabla}\rho \cdot {\bf \nabla} f_k}{f} - 9\rho \frac{\left({\bf \nabla} f\right)^2}{f} \right]
\frac{j_1(k_F s)}{k_F s}\right\}\nonumber \\
&&- \frac{m^2}{\hbar^4}\frac{\rho W^2}{f^2}s^2 \frac{j_1(k_F s)}{k_F s} -
\frac{im}{2\hbar^2}\frac{\rho}{f}\hat{\sigma}\cdot\left({\bf W}\times{\bf s}\right)
\frac{3j_1(k_F s)}{k_F s}\Bigg]_{k=k_F} \, ,
\label{eq7}
\end{eqnarray}
\end{widetext}
where $k_F$ is the local Fermi momentum.
The quantity $f\left({\bf R},k\right)$  in Eq.~(\ref{eq7}) is the inverse of the position- and momentum- dependent effective mass,
which at ETF-$\hbar^2$ level is defined as:
\begin{equation}
f\left({\bf R},k\right) = 1 + \frac{m}{\hbar^2k}\frac{\partial V^{nucl}_{exch,0}({\bf R},k)}{\partial k}.
\label{eq:effmass}
\end{equation}
In this equation  $V^{nucl}_{exch,0}$ is the Wigner transform of the exchange potential at $\hbar^0$ (TF level) for nucleons of a given type,
which depends of both, neutron and proton, densities (see Eq.(9) of~\cite{bha21} for the explicit expression of $V^{nucl}_{exch,0}$). 
In Eq.~(\ref{eq7}) $f_k$ and $f_{kk}$ stand for the first and second derivatives of $f\left({\bf R},k\right)$ with respect to $k$, while $\nabla f$ 
and $\nabla f_k$  correspond to its spatial derivatives. All these derivatives are evaluated at $k=k_F$.
The last two terms of Eq.~(\ref{eq7}) are the contributions due to the spin-orbit force, which depend of the form-factor of the spin-orbit potential
$\bf W$  for each type of nucleons (see Eq.(10) of~\cite{bha21}). 

This ETF formalism is completely general for interactions with a finite-range central term as for example the Gogny interaction considered in this 
work. The impact of the finite-range part of the interaction on the density matrix appears encoded in the inverse effective mass defined in~(\ref{eq:effmass})
and in its momentum and spatial derivatives. 
A comparison of the SV density matrix with the earlier expansions NV and CB can be found in Ref.~\cite{sou00}.

The explicit form of the semiclassical kinetic energy and spin densities at ETF level can be easily derived from the density matrix~(\ref{eq7}) as:
\begin{eqnarray}
\tau_{SV} \left({\bf R}\right) &=& \left. \left(\frac{1}{4}\Delta_R - \Delta_s \right)
{\tilde \rho}_{SV}\left({\bf R},s\right)\right\vert_{s=0} \nonumber\\
 &=& \frac{3}{5} k_F^2 \rho
+ \frac{1}{12}\Delta \rho \left[4 + \frac{2}{3}k_F \frac{f_k}{f} \right]
+\frac{1}{36}\frac{\left({\bf \nabla} \rho\right)^2}{\rho} \left[ 1 + \frac{2}{3}k_F \frac{f_k}{f} +
\frac{2}{3}k_F^2 \frac{f_{kk}}{f}- \frac{1}{3}k_F^2 \frac{f_k^2}{f^2} \right]
\nonumber \\
&+&\frac{1}{6}\frac{\rho}{f} \left[\Delta f - \frac{({\bf \nabla}f)^2}{2f} \right]
+ \frac{1}{6}\frac{{\bf \nabla}\rho \cdot {\bf \nabla}f}{f}
\left[1 - \frac{1}{3}k_F \frac{f_k}{f}\right]
+ \frac{1}{9}\frac{{\bf \nabla}\rho \cdot {\bf \nabla}f_k}{f}
+ \frac{1}{2}\left(\frac{2m}{\hbar^2}\right)^2 \frac{\rho}{f^2} W^2.
\label{eq:tau2*}
\end{eqnarray}
and the spin density at ETF level can be expressed as: 
\begin{eqnarray}
{\bf J}_{SV} 
&=& -i \left[\sigma \times \left(\frac{{\bf \nabla}_R}{2} + {\bf \nabla}_s)\right)\right]
\left. {\tilde \rho}_{SV}\left({\bf R},s\right)\right\vert_{s=0}
= - \frac{2m}{\hbar^2}\frac{\rho {\bf W}}{f},
\label{eq9}
\end{eqnarray}
which enters in the kinetic energy density (\ref{eq:tau2*})  and multiplied by the form-factor of the spin-orbit potential $\bf W$
provides the spin-orbit energy for each type of nucleons.

Therefore at ETF level, the kinetic energy and spin densities become functionals of the particle densities for neutron and protons. 
If the effective mass is independent of the momentum, as for the zero-range Skyrme forces, the kinetic energy reduces to the $\hbar^2$ expansion 
reported in Ref.~\cite{bra85} whereas, if the effective mass  is equal to the physical one, the kinetic energy density reduces to the well-known 
Weizs\"acker term.

The semiclassical approximation of the density matrix has also an important impact on the exchange energy density, which is now local and for each 
kind of nucleon can be written as~\cite{sou00,bha21}
The HF exchange energy density is given by
\begin{equation}
{\cal H}_{HF}^{exch} = \frac{1}{2} \int V_ {exch}^{nucl}({\bf R},{\bf s})
\rho({\bf R},{\bf s}) \,
d{\bf R} \,d{\bf s} \, ,
\label{eq22}
\end{equation}
where the exchange potential is defined as $V_ {exch}^{nucl}({\bf R},{\bf s})= - v({\bf R},{\bf s}) \rho({\bf R},{\bf s})$ 
If the quantal density matrix in~(\ref{eq22}) is replaced by the semiclassical one~(\ref{eq7}), the exchange energy can be written locally as the sum of a $\hbar^0$ (Slater) contribution, which corresponds to
the exchange energy in infinite nuclear matter, plus a $\hbar^2$ contribution, which can be finally
recast in terms of the neutron and proton densities and their gradients up to second order~\cite{sou00,bha21}:
\begin{eqnarray}
{\cal H}^{nucl} _{exch}({\bf R}) &=& {\cal H}^{nucl}_{exch,0}({\bf R}) + {\cal H}^{nucl}_{exch,2}({\bf R}),
\end{eqnarray}
where the $\hbar^2$ contribution to the exchange energy reads:
\begin{eqnarray}
{\cal H}^{nucl}_{exch,2}({\bf R}) 
&=&
\frac{\hbar^2}{2m} \left[ (f-1)  \left( \tau - \frac{3}{5} \rho k_F^2 - \frac{1}{4} \Delta \rho \right)
+ k_F f_k \left(\frac{1}{27} \frac{(\nabla \rho)^2}{\rho} - \frac{1}{36} \Delta \rho \right) \right]
\end{eqnarray}

\subsection{Results}

We have performed calculations of the surface energy in semi-infinite nuclear matter using the SV approach described previously, which
 includes explicitly contributions of the momentum-dependent effective mass and of the spin-orbit interaction, in contrast to the previous 
approximations discussed above. Although the minimization of the surface energy~(\ref{eq:surf}) at ETF level could be done by solving the
Euler-Lagrange corresponding to neutron and proton densities~\cite{bha21}, to be consistent with the previous calculations we perform a restricted
variational calculation using the trial density $\rho(z)$ (\ref{density:prof:SINM}). 
The resulting surface energy coefficients are displayed in  Table~\ref{table:nonlocal} for the first group
interactions, and the four cases: with and without spin--orbit term and with and without an effective mass.
  
\begin{table}[!h]
\begin{center}
\begin{tabular}{|l|cc|cc|}
\hline
& \multicolumn{2}{|c|}{SV} & \multicolumn{2}{|c|}{SV ($f=1$)}  \\
\hline
& no SO & SO & no SO & SO  \\
\hline
D1      & 20.24 & 17.60 & 21.53 & 19.07 \\
        & 2.2399 & 2.2155 & 2.0541 & 2.4506 \\
\hline
D1S    & 18.60  & 15.57 & 20.05  & 17.14 \\
       & 2.1718 & 2.7876 & 1.9784 & 2.4068 \\
\hline
D1N    & 18.54  & 16.18 & 19.49 & 17.18 \\
       & 2.2255 & 2.6460 & 2.1008 & 2.4455 \\
\hline
D1M    & 18.74 & 16.14 & 19.77 & 17.23 \\ 
       & 2.2771 & 2.8011 & 2.1312 & 2.5691 \\
\hline
D1M*  & 18.98 & 16.31 & 19.82 & 17.30 \\
       & 2.2923 & 2.8345 & 2.1317 & 2.5384 \\
\hline
D1MK & 18.80 & 16.20 & 19.85 & 17.36 \\  
      & 2.2771 & 2.8016 & 2.1641 & 2.5740 \\
\hline
D3G3 & 18.70 & - & 19.32 & - \\ 
      & 2.3597 & - & 2.2270 & - \\
\hline
D3G3M & 19.61 & - & 19.83 & - \\  
       & 2.3049 & -& 2.2413 & - \\
\hline
\end{tabular}
\caption{Surface energy coefficients (in MeV) obtained using the SV approximation including (or not) the spin-orbit term and including (or not) 
the momentum-dependent effective mass. For each interaction, we also provide the value of the variational parameter $\alpha$ (in fm$^{-1}$). See text for details.}
\label{table:nonlocal}
\end{center}
\end{table}

In the absence of an explicit spin--orbit contribution, the SV approximation always provides a variational minimum for all interactions 
belonging to the first group. The quality of the results depends on the specific interaction, but, in general, the SV approximation reproduces 
the HF values within an accuracy of approximately 0.5 MeV. For the interactions belonging to the second group, the discrepancy is larger, 
reaching values of about 1 MeV. In the D280, it was not possible to find a minimum within a reasonable range.

The inclusion of the spin--orbit interaction further deteriorates the agreement with the HF results, increasing the discrepancy to approximately 
1.5--2 MeV. Moreover, the inclusion of the spin--orbit term does not always lead to a minimum of $a_s(\alpha)$.

In the last two columns are displayed the results by setting the effective mass equal to its bare value, \emph{i.e.} by imposing $f=1$. In the 
absence of the spin--orbit contribution, this approximation tends to overestimate the HF results by at most 0.6 MeV for the interactions belonging 
to the first group. For the interactions of the second group, the deviations are significantly larger, ranging from an underestimation of the 
D280 HF value by 170 keV to an overestimation of the D300 HF result by 5.2 MeV.
This shows that, for the Gogny interactions considered here, non-local effects have a much stronger impact than in the Skyrme case, where their 
contribution can be estimated to be of the order of 1 MeV.

When the spin--orbit contribution is included, only a subset of the interactions belonging to the first group leads to a variational minimum. 
In these cases, the semiclassical approximation systematically underestimates the HF results, but the discrepancy remains typically within 
about 1 MeV.
 

\section{Conclusions}\label{sec:conclusion}

In this work, we have investigated several semiclassical approximations for the exchange contribution of Gogny interactions to calculate the surface energy coefficient in semi-infinite nuclear matter.

By comparing the different semiclassical schemes with fully self-consistent Hartree--Fock calculations performed for the same system, we have shown that, in the absence of the spin--orbit interaction, second-order extended Thomas--Fermi (ETF2) calculations combined with either the Slater, Negele--Vautherin, or Campi--Bouyssy density-matrix expansions provide surface energy coefficients in good agreement with the HF values, with typical deviations of only a few hundred keV.

The inclusion of the spin--orbit interaction introduces additional difficulties. In this case, the most accurate results are obtained when the ETF4 approximation is adopted for the kinetic-energy density, combined with either the SL/NV or CB prescriptions for the density matrix. Nevertheless, a systematic correction of a few hundred keV remains necessary to recover the HF results accurately.

For a limited number of interactions, namely D260, D280, and D300, and to a lesser extent D3G3M, the semiclassical expansion exhibits a significant loss of accuracy and may eventually fail to provide a meaningful estimate of the surface coefficient. This behaviour is associated with the appearance of enhanced Friedel oscillations, which become particularly pronounced when the spin--orbit contribution is included. We have found that these interactions are characterized by a larger relative contribution of the exchange term to the total binding energy and by significantly higher values of the nuclear incompressibility. Although this observation provides a useful criterion to identify problematic interactions within the present data set, a dedicated study based on newly adjusted Gogny parametrizations would be required to disentangle the individual role of these different ingredients.

We have explored the influence of non-local effects through the introduction of a momentum-dependent effective mass. The present analysis, restricted to the ETF2 level, shows that these contributions have a significantly larger impact on the surface energy coefficient for finite-range Gogny interactions than for zero-range Skyrme forces. This highlights the importance of properly accounting for non-local effects when developing semiclassical approximations for finite-range effective interactions.

Our primary objective is to provide a reliable estimate of the surface energy coefficient at a reduced computational cost, in order to be directly employed during an optimisation procedure of the parameters of the interaction. In this way, one can determine whether a given parametrization is expected to yield a trustworthy estimate of the surface energy coefficient without performing a full HF calculation at every iteration of the optimization procedure. We have shown that a local ETF4 approximation fulfill this requirement. In Appendix~\ref{app:B}, we provide a simple pocket formula based on the fourth-order Thomas--Fermi expansion within the Slater approximation. Exploiting the approximately quadratic dependence of the surface energy coefficient on the $\alpha$ parameter, we propose evaluating the formula at three selected points and using a parabolic interpolation to locate the minimum. The corresponding coefficient can then be determined with an accuracy of approximately 200--300~keV, which is sufficient for its inclusion in a fitting procedure.

\section*{Acknowledgment}

We thank K. Bennaceur for providing us the numerical results in finite nuclei for various Gogny interactions. 
One of us, XV, acknowledges financial support from Grants Nos.\ PID2023-147112NB-C22 and CEX2024-001451-M funded by the Spanish 
MICIU/AEI/10.13039/501100011033.

\appendix 
\section{Analytical formulae}\label{app:A}

We provide here the complete expression of the $G(x),H(x)$ functions entering Eq.\eqref{eq:VE}-\eqref{eq:NV:Hx}

\begin{equation}
G(x) = {\rm e}^{-x^2}\left[ \frac{1}{x} - \frac{2}{x^3} \right]  +  \frac{2}{x^3} - \frac{3}{x} + \sqrt{\pi} \, {\rm Erf}(x).
\end{equation}

\begin{equation}
H(x) = \frac{5 \sqrt{\pi } x^3
   \text{Erf}(x)-2 \left(x^4+6 x^2-4\right)+4 {\rm e}^{-x^2}
   \left(x^2-2\right)}{x^5}.
\end{equation}

We also provide the expression of $F(x)$ entering Eq.\eqref{eq:NV:Fx} at various levels of approximation

\begin{eqnarray}
F_{TF}(z) &=& \frac{1}{4} \alpha^2 x(z)^2 (x(z)-1)(2x(z)-1) \\
F_{ETF2}(z) &=& - \frac{1}{36} \alpha^2 x^2 (x(z)-1)(7x(z)-4) \\
F_{ETF4}(z) &=& - \frac{1}{36} \alpha^2 x^2 (x(z)-1)(7x(z)-4) - \left(\frac{3\pi^2}{2}\right)^{-2/3}\frac{1}{540} \alpha^4 \rho^{1/3} (-2+21x-39x^2+20x^3)
\end{eqnarray}

The expressions of $\hat k$ for the CB approximation defined in Eq.~\eqref{eqn:CB} at different orders
\begin{eqnarray}
{\hat k}^2_{TF} &=& \left(\frac{3 \pi^2}{2} \rho \right)^{2/3} - \frac{5}{12} \alpha^2 (x(z)-1)(2x(z)-1) \\
{\hat k}^2_{ETF2} &=& \left(\frac{3 \pi^2}{2} \rho \right)^{2/3} + \frac{5}{108} \alpha^2 (x(z)-1)(7x(z)-4) \\
{\hat k}^2_{ETF4} &=& \left(\frac{3 \pi^2}{2} \rho \right)^{2/3} + \frac{5}{108} \alpha^2 (x(z)-1)(7x(z)-4) + \alpha^4 \frac{x(z)^{1/3}}{324 (3 \pi^2 \rho_0)^{2/3}} (x(z)-1)(2-19 x(z)+20 x(z)^2)
\end{eqnarray}

\section{A practical formula for a fit}\label{app:B}
We collect here the previous results, in a very compact form, to get the ETF4 semiclassical surface energy for a standard Gogny interaction, neglecting non-local effects.
\begin{eqnarray}
\frac{E}{S} &=& A \alpha^3 + B \alpha + \frac{C}{\alpha} - \frac{3}{8} \, t_3 \,  \rho_0^{\gamma+2} C_\gamma \frac{1}{\alpha} \nonumber \\
&& -\frac{3}{64}  W_0^2 \rho_0^3 \frac{m}{\hbar^2} \alpha + \left(\frac{3\pi^2}{2}\right)^{-2/3}\frac{\hbar^2}{2m} \frac{9W_0^2}{16} \left(\frac{m}{\hbar^2}\right)^2 \rho_0^{7/3} \left\{ \frac{9}{728}
+ \frac{2187}{695552} W_0^2 \left(\frac{m}{\hbar^2}\right)^2 \rho_0^3 
\right\} \alpha^3 \nonumber \\
&& + \frac{1}{8} \rho_0^2 
\sum_i ( 4W_i + 2B_i  - 2 H_i - M_i ) (\sqrt{\pi} \mu_i)^2 \int_{-\infty}^{\infty} {\rm d}z \, x(z) 
\left[ \left( \int_{-\infty}^{\infty} {\rm d}z' \, {\rm e}^{-(z - z ')^2 / \mu_i^2} x(z') \right) - \sqrt{\pi} \mu_i \right] \nonumber \\
&& 
- \frac{\rho_0}{2 \sqrt{\pi}}   \sum_i \left( W_i + 2B_i  - 2 H_i - 4 M_i \right) 
\int_{-\infty}^{\infty} {\rm d}z \, x(z) 
\left[ G(k_F(z) \mu_i) -  G(k_{F0} \mu_i) \right].
\end{eqnarray}
where $x(z)=(1+e^{\alpha z})^{-1}$, 
the expressions of $A, B, C$ are given in Eq.\eqref{Energy1}-\eqref{Energy3}, the coefficient $C_{\gamma}$ in Table \ref{tab:dd_power} and  the function $G$ in Appendix~\ref{app:A}.  
Based on the results presented in Fig.~\ref{fig:alpha:SO}, the full minimisation with respect to the variational parameter $\alpha$ can be avoided by evaluating the function at three values of chosen around the average value $\alpha = 2.13~\mathrm{fm}^{-1}$. A parabolic interpolation of these points will determine its minimum and the surface energy coefficient, providing an efficient and computationally inexpensive alternative to a direct minimisation procedure.

\bibliography{biblio}

\end{document}